\documentclass[aps,prb,twocolumn,groupedaddress]{revtex4}

\usepackage{amsmath}
\usepackage{epsfig}
\usepackage{graphics}

\newcommand{\teff}{T_{\rm eff}}
\newcommand{\hdet}{H_{\rm det}}
\newcommand{\henv}{H_{\rm env}}
\newcommand{\henvcoup}{H_{\rm ecoup}}
\newcommand{\hnoise}{H_{\rm noise}}
\newcommand{\hcoup}{H_{\rm coup}}

\newcommand{\noise}{S_I}
\newcommand{\third}{\delta^3 I}
\newcommand{\Tav}{\bar{T}_{\rm TLS}}
\newcommand{\dT}{\Delta T_{\rm TLS}}

\renewcommand{\Re}{{\rm Re}}

\newcommand{\sgn}{{\rm sgn}}

\newcommand{\GN}{G}

\def\ba{\begin{array}}
\def\ea{\end{array}}

\begin{document}

\title{Quantum detectors for the third cumulant of current
fluctuations}

\author{Tero T. Heikkil\"a}
\email[Correspondence to ]{Tero.Heikkila@tkk.fi}

\author{Teemu Ojanen}

\affiliation{Low Temperature Laboratory, P.O. Box 3500, FIN-02015
TKK, Finland}

\date{\today}

\begin{abstract}
We consider the measurement of the third cumulant of current
fluctuations arising from a point contact, employing the transitions
that they cause in a quantum detector connected to the contact. We
detail two generic detectors: a quantum two-level system and a
harmonic oscillator. In these systems, for an arbitrary relation
between the voltage driving the point contact and the energy scales
of the detectors, the results can be expressed in terms of an
effective detector temperature $\teff$. The third cumulant can be
found from the dependence of $\teff$ on the sign of the driving
voltage. We find that proper ordering of the fluctuation operators
is relevant in the analysis of the transition rates. This is
reflected in the effective Fano factor for the third cumulant
measured in such setups: it depends on the ratio of the voltage and
an energy scale describing the circuit where the fluctuations are
produced.
\end{abstract}


\maketitle

\section{Introduction}
\label{sect:intro}

The statistics of current fluctuations in mesoscopic conductors have
been at the center of interest within the last decade or so. This
statistics can be described by the characteristic function, which is
the Fourier transform of the probability density for a given value
of the current. Characteristic functions for many different types of
systems have been calculated, ranging from the simple case of a
tunnel junction with Poisson distributed currents, to point contacts
with binomial distribution and to more complicated systems composed
for example of superconductors.\cite{belzig03,nazarovbook} Along
with the characteristic function, the fluctuations can be described
by the cumulants or moments of the distribution, the latter being of
the form
\begin{equation}
M_n(t_1,\dots,t_n)=\langle \prod_{i=1,n}\delta I(t_i) \rangle,
\label{eq:clmoment}
\end{equation}
where $\delta I(t_i) = I(t_i)-\langle I \rangle$, $I(t)$ is the
instantaneous value of the current, and the brackets $\langle \cdot
\rangle$ refer to ensemble averaging. Typically described observable
is then the Fourier transform of $M_n(0,t_2-t_1,\dots t_n-t_1)$ with
respect to the time differences $t_i-t_1$, and often only the limit
where all the frequencies are taken to zero is known.

As many of the studied systems require quantum mechanics for their
description, a natural question is the proper generalization of
Eq.~\eqref{eq:clmoment} to include the fact that the current
operator $\hat{I}(t)$ may not commute with itself at different
times. The answer to this question depends on how the fluctuations
are to be measured. Levitov, Lee and Lesovik suggested the use of a
spin coupled to the fluctuating current as an imaginary
detector.\cite{levitov:4845} More precisely, when the coupling is
turned on at time $t=0$, the angle of the spin coupled only to the
fluctuating current $\hat{I}(t)$, with no average fields, starts to
precess along with the current. Then the angle at a later time $t$,
averaged over the fluctuations, is
\begin{equation}
\begin{split}
&\sigma_+(t) = \sigma_+(0)
\\&\times\left\langle \tilde{T} \exp\left(i\frac{\tilde{g}}{2}\int_t^0 dt
\hat{I}(t) \right) T \exp\left(-i\frac{\tilde{g}}{2}\int_0^t dt
\hat{I}(t)\right)\right\rangle . \label{eq:spindetector}
\end{split}
\end{equation}
Here $T$ and $\tilde{T}$ denote the time- and anti-time-ordering
operators and $\tilde{g}$ is the coupling constant. This combination
of time-ordering operators is also called the Keldysh ordering. For
a classical current, ignoring noncommutativity,
Eq.~\eqref{eq:spindetector} would yield the characteristic function
of the charge $Q=\int_0^t I(t)$ transmitted in a conductor in a
given time $t$. Hence, Eq.~\eqref{eq:spindetector} is one possible
generalization of the characteristic function for a quantum current
operator. In the language of quantum two-level systems (TLS's) or
qubits, Eq.~\eqref{eq:spindetector} describes the dephasing of a
non-biased (zero level spacing) qubit.

The advantage of the spin detector is the fact that the
Keldysh-symmetrized characteristic function of a point contact with
transmission probability $T_n$ for channel $n$ is a product of
binomial characteristic functions, described by the probability
$T_n$ of success. For example, the first three moments of the
distribution of currents measured in that way are
\begin{equation*}
\begin{split}
\langle I \rangle &= G_0 V \sum_n T_n\\
\langle \delta I^2 \rangle &= G_0 e V \sum_n T_n(1-T_n) \equiv F_2 e
\langle I \rangle\\
\langle \delta I^3 \rangle &= G_0 e^2 V \sum_n T_n(1-T_n)(1-2T_n)
\equiv F_3 \langle I \rangle.
\end{split}
\end{equation*}
Here $G_0=e^2/h$ and the summation goes over the spin and the
different transverse channels in the point contact. Such a
characteristic function thus has a transparent classical
interpretation.

But spin precession or qubit dephasing in real time is difficult to
measure. Also, Levitov-Lee-Lesovik spin detector responds only to
the zero-frequency noise. An alternative method is to measure the
excitation and relaxation rates of a quantum system subject to a
fluctuating force produced by the current. This is the approach
taken in the present paper. These transition rates determine the
static state of the density matrix for the system, which can in many
cases be described by an effective temperature. Measuring the
effective temperature is much more simple than spin precession, and
it also gives access to the finite-frequency moments of the noise.
In this paper we discuss how two simple quantum detectors behave in
response to a non-Gaussian fluctuating force. An exact solution of
this problem with the full characteristic function of fluctuations
is not known to us, and therefore we resort to expanding in the
cumulants of these fluctuations up to the third cumulant.

It turns out that in this case the proper way to define the
third-order correlation function is by time-ordering two of the
current fluctuation operators, and leaving the third one free (c.f.,
Eq.~\eqref{eq:third}).\cite{ojanen:020501} The relaxation of the
qubit in this case is sensitive to the frequency dependence of the
third cumulant,\cite{saloup06} and the relevant Fano factor
"measured" in this process depends on the relation between different
frequency scales of the problem: If the qubit is "close" to the
noise source, so that the frequency dependence is given by the
voltage $V$ applied over the scatterer, the third cumulant strength
is characterized by $F_2-F_3=2\sum_n T_n^2 (1-T_n)/\sum_n T_n$. This
factor is very small for a tunnel junction, but finite for other
types of scatterers. However, placing the qubit further from the
noise source introduces another frequency scale into the problem,
characterizing the circuit between the source and the qubit. For
voltages above this frequency scale, the third cumulant effect on
relaxation is characterized by the "usual" Fano factor, $F_3$. In
this limit one reproduces our results in
Ref.~\onlinecite{ojanen:020501}, where we assumed that the frequency
dependence is solely governed by the circuit. Besides the Fano
factor, also the dependence on the voltage $V$ is different in the
two limits: in the first case it is logarithmic and in the second
case linear.

Alternative "on-chip" detector schemes for measuring non-Gaussian
fluctuations have been suggested in
Refs.~\onlinecite{tobiska:106801,sonin:140506,heikkila:247005,pekola:206601,brosco:024524,ankerhold:186601,ankerholdup06}
and a few schemes have already been experimentally realized, see
Refs.~\onlinecite{lindell:197002,gustavsson:076605,gustavssonup06a,gustavssonup06b}.

\subsection{Frequency scales and different regimes}

The noise source - detector system can be characterized with a few
frequency/energy scales whose relative magnitudes determine the
detector response. The main frequency scales are that coming from
the detector level spacing $\Omega/\hbar$, one characterizing the
circuit connecting the noise source and the detector, say $\omega_c$
(for an example, see Ref.~\onlinecite{ojanen:020501}), and the rate
$\Gamma_{\rm env}$ of transitions in the detector induced by its own
environment. Furthermore, the fluctuations in a point contact are
characterized by two further scales, given by the temperature $T$
and the voltage $V$ over the contact.

In this work, we assume that the level broadening is sufficiently
weak, i.e., $\hbar \Gamma_{\rm env} \ll \Omega$, such that it can be
taken into account perturbatively. The detector response will depend
on the ratios between the voltage and the detector level spacing,
$eV/\Omega$ (or, at a finite temperature, on $k_B T/\Omega$) and on
the ratio $\hbar \omega_c/\Omega$. Finally, the response depends on
the relative magnitude of the transition rates coming from the noise
source and of those coming from the Gaussian bath.

\begin{figure}[h]
\centering
\includegraphics[width=\columnwidth]{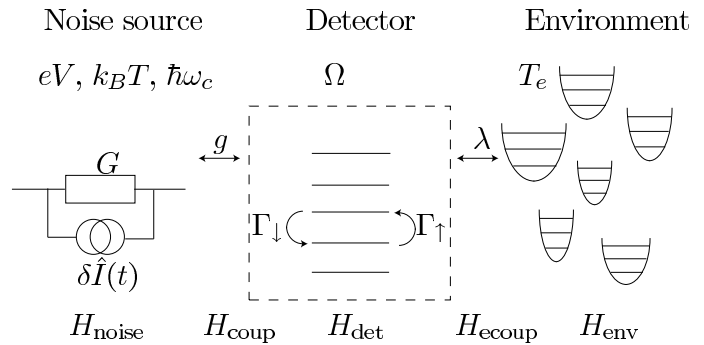}
\caption{Schematic idea of the fluctuation measurements: Current
fluctuations in the noise source are coupled to a quantum detector,
where they induce transitions between the detector energy levels.
Another source for the transition rates is the intrinsic Gaussian
environment of the detector, which can be modeled via an ensemble of
harmonic oscillators. Up to the third order in the coupling
coefficient $g$, the transition rates in the detector depend on the
second and third cumulant of the fluctuations. As the latter is odd
in the driving average current through the noise source, its effect
can be measured by detecting the change in the rates when the sign
of the driving average current is reversed.} \label{fig:setup}
\end{figure}

\section{Effect of third cumulant on an arbitrary detector}
Consider the system depicted in Fig.~\ref{fig:setup}. A non-Gaussian
noise source is coupled to a detector whose state we aim to
describe. The Hamiltonian of the system can be decoupled into
\begin{equation*}
H=\hdet+\henv + \henvcoup + \hnoise + \hcoup.
\end{equation*}
Here $\hdet$ is the Hamiltonian of the detector, specified in
Secs.~\ref{sec:tls} and \ref{sec:ho}. It is in general described
through a set of collective variables with mutually noncommuting
operators $\hat{A}_i$. This detector is coupled to its own
environment described by $\henv$ through the coupling $\henvcoup$.
We assume this environment to be Gaussian, such that we can model it
via an ensemble of harmonic oscillators,\cite{caldeira:374}
\begin{equation*}
\begin{split}
  \henv &= \sum_{j} \hbar \omega_j \hat{a}_j^\dagger \hat{a}_j, \\
  \henvcoup &= \hat{A} \sum_j \lambda_j (\hat{a}_j^\dagger +
  \hat{a}_j) + \frac{1}{\hbar\omega_0}\hat{A}^2 \sum_j \lambda_j^2,
  \end{split}
\end{equation*}
where $\hat{a}^{(\dagger)}$ is the bosonic annihilation (creation)
operator of an oscillator in the bath, $\omega_n$ its
eigenfrequency, and $\lambda_j$ is the coupling constant from this
oscillator to the system variable $\hat{A}$.

Finally, the noise source is described by $\hnoise$ and coupled to
the detector via $\hcoup$. The latter connects in general a set of
collective variables of the noise source to another set of variables
of the detector. As we aim to describe the measurement of current
fluctuations, we explicitly assume that the previous is the current
operator $\hat{I}$ in the noise source.\cite{voltagenote} Thus the
coupling is of the form
\begin{equation*}
\hcoup = \frac{\hbar}{e}\sum_i g_i \hat{I} \hat{A}_i,
\end{equation*}
where $g_i$ are dimensionless coupling constants. We further assume
that $\hat{I}$ commutes with $\hat{A}_i$.

Including the effects of $\hcoup$ up to the third order in the
coupling $g_i$, we do not have to specify $\hnoise$, but it suffices
to concentrate on the different correlators (cumulants) of
$\hat{I}$. For simplicity, we include the effect of the average
current on the detector Hamiltonian $\hdet$, such that we can take
$\langle \hat{I} \rangle =0$. Here the brackets $\langle \cdot
\rangle$ denote quantum averaging over the density matrix of the
noise source. In particular, we concentrate on the noise power
spectral density,
\begin{equation*}
\noise(\omega) \equiv \int_{-\infty}^\infty dt e^{i \omega (t-t_0)}
\langle \hat{I}(t) \hat{I}(t_0) \rangle,
\end{equation*}
and the partially time-ordered third
cumulant\cite{ojanen:020501,fouriernote}
\begin{equation}
\begin{split}
\third(\omega_1,\omega_2) \equiv &\int_{-\infty}^\infty d (t_1-t_0)
d(t_2-t_0) e^{i \omega_1(t_1-t_0)}\\&\times e^{i \omega_2(t_2-t_0)}
\langle \tilde{T}[\hat{I}(t_0)\hat{I}(t_1)] \hat{I}(t_2)\rangle.
\end{split}
\label{eq:third}
\end{equation}
In a static system considered in this paper, these correlators are
independent of the time $t_0$.

Assume the uncoupled detector is described via the energy
eigenstates $|n\rangle$ with energies $E_n$, i.e., $\hdet |n\rangle
= E_n |n \rangle$. Without coupling to the environment, the state of
the detector is described by a density matrix $\rho_{nm}^{\rm
det}(t)$ which in general may show coherent oscillations between the
different states. Assume now we turn on the couplings $\hcoup$ and
$\henvcoup$ at some time $t_0$. If $\hcoup$ or $\henvcoup$ do not
commute with $\hdet$, after some time $\rho_{nm}^{\rm det}(t)$ tends
into a diagonal steady state form.\cite{brosco:024524} These
diagonal entries $P_n \equiv \rho_{nn}^{\rm det}(t \gg t_0)$ are
obtained from a detailed-balance relation of the form
\begin{equation}
\frac{P_n}{P_m} = \frac{\Gamma_{m\rightarrow n}}{\Gamma_{n
\rightarrow m}}. \label{eq:detailedbalance}
\end{equation}
In addition, the total probability has to be conserved, i.e.,
$\sum_n P_n = 1$. Here $\Gamma_{m \rightarrow n}$ is the total
transition rate from the energy eigenstate $m$ to the eigenstate
$n$, due to the coupling to the environment.

Up to the third order in the coupling constants $g_i$ and
$\lambda_i$, the transition rates originating from the coupling to
the oscillator bath and the noise source are uncorrelated and we may
write
\begin{equation*}
\Gamma_{m \rightarrow n} = \Gamma_{m \rightarrow n}^{\rm env} +
\Gamma_{m \rightarrow n}^{\rm noise}.
\end{equation*}
Here
\begin{equation*}
\Gamma_{m \rightarrow n}^{\rm env} =  \frac{|A^{mn}|^2}{1-e^{-\hbar
\omega_{mn}/(k T_e)}} \sum_j \lambda_j^2 = e^{\hbar \omega_{mn}/(k
T_e)} \Gamma_{n \rightarrow m}^{\rm env}
\end{equation*}
is the transition rate from state $m$ to state $n$ due to the
coupling of the detector to its Gaussian bath with temperature
$T_e$, up to the second order in the coupling constants $\lambda_j$.
Here $A^{mn} \equiv \langle m | \hat{A} | n \rangle$ and
$\omega_{mn} \equiv (E_m-E_n)/\hbar$. In what follows, we use a
short-hand notation $\lambda^2 \equiv \sum_j \lambda_j^2$. The
transition rates induced by the noise source are of the form
\cite{schoelkopf02,ojanen:020501}
\begin{equation*}
\Gamma_{m \rightarrow n}^{\rm noise} = \frac{1}{e^2} \sum_i g_i^2
|A_i^{mn}|^2 \noise(\omega_{mn}) + \Gamma_{m \rightarrow n}^{(3)}.
\end{equation*}
The rate from the third-order term is\cite{ojanen:020501}
\begin{equation}
  \Gamma_{m\rightarrow n}^{(3)} = \sum_i \frac{g_i^3}{e^3} \Re
  \sum_{l} \left[\int d\omega
  \frac{\third(\omega,-\omega_{mn})}{\omega-\omega_{ln}-i\eta}
  A_i^{ml} A_i^{ln} A_i^{nm}\right].
  \label{eq:generalthirdorder}
\end{equation}
Here $\eta$ is a positive infinitesimal. In this paper, this
integral is evaluated for two generic detectors in the case of noise
originating from a point contact.

\section{Frequency dependent second and third cumulants of a point contact}
A general scheme for calculating arbitrarily ordered
frequency-dependent current correlators from the scattering theory
was laid out by Salo, Hekking and Pekola (SHP) in
Ref.~\onlinecite{saloup06}. Their results are applied here in order
to calculate the response of our generic detectors to the
third-order current fluctuations. SHP decompose the operator
describing the current through a given scatterer to "in" and "out"
parts, $\hat{I}=\hat{I}_{\rm in}-\hat{I}_{\rm out}$, and show that
time ordering between two current operators, $\hat{I}(t_1)$ and
$\hat{I}(t_2)$, can be expressed in terms of ordering between
$\hat{I}_{\rm in}$ and $\hat{I}_{\rm out}$. For the latter, they
find that a time-ordered product of a pair of (in,in) or (out,out)
operators is the same as the unordered pair, and that time ordering
a pair of (in,out) operators corresponds to an ordering where
$\hat{I}_{\rm out}$ is placed to the left of $\hat{I}_{\rm in}$.

A practically important noise source is a point contact with an
energy independent scattering matrix. A "point contact" in this case
refers to a system through which the electron time-of-flight
$\tau_D$ is much smaller than the other time scales of the problem.
In this case, the second-order correlator can be written in the form
$\noise(\omega)=\noise^Q(\omega)+\noise^{\rm exc}(\omega)$, where
the vacuum fluctuations are $\noise^Q(\omega) = 2 \hbar \omega \GN
\theta(\omega)$, and the excess noise is given by
\cite{aguado:1986,heikkila:247005}
\begin{widetext}
\begin{equation*}
\begin{split}
\noise^{\rm exc}(\omega)&=\GN \hbar
\omega\left(\coth\left(\frac{\hbar \omega}{2 k
T}\right)-\sgn(\omega)\right) + F_2 \GN
\frac{eV\sinh(\frac{eV}{kT})-2\hbar \omega \coth(\frac{\hbar
\omega}{2kT}) \sinh^2(\frac{eV}{2kT})}{\cosh(\frac{eV}{kT})-\cosh(\frac{\hbar \omega}{kT})}\\
&\overset{T\rightarrow 0}{\rightarrow} F_2 \GN (e|V|-\hbar
|\omega|)\theta(e|V|-\hbar |\omega|).
\end{split}
\end{equation*}
\end{widetext}
Here $\GN$ is the conductance of the point contact, $V$ is the
voltage applied over it, and $F_2 = [\sum_n T_n (1-T_n)]/\sum_n T_n$
is the Fano factor characterizing the transmission eigenvalues $T_n$
of the contact. Written in this way, the excess noise is a symmetric
function of frequency, and thus it contributes to excitation as much
as to relaxation.

For the partially time-ordered third cumulant in the case of a point
contact, one can deduce from the results of SHP\cite{shpfouriernote}
\begin{equation}
\begin{split}
&\third(\omega_1,\omega_2) =
S_{ioo}(-\omega_1-\omega_2,-\omega_2)+S_{ioo}(\omega_1,-\omega_2)\\&+
S_{ioo}(-\omega_2,-\omega_1-\omega_2)-S_{ooo}(-\omega_1-\omega_2,-\omega_2).
\end{split}
\label{eq:pcthird}
\end{equation}
Here
\begin{align*}
&S_{ioo}(\omega_1,\omega_2)=eF_2 G
[A(\omega_1,\omega_2-v)-B(\omega_1,\omega_1-\omega_2-v)]\\
\begin{split}
&S_{ooo}=eF_3 G
[A(\omega_1,\omega_2-v)+A(\omega_1-v,\omega_2)\\&+A(\omega_1+v,\omega_2+v)-
B(\omega_1,\omega_1-\omega_2-v)\\&-B(\omega_1-v,\omega_1-\omega_2)-B(\omega_1+v,\omega_1-\omega_2+v)],
\end{split}
\end{align*}
$v=eV/\hbar$ and $F_3 =  \sum_n T_n(1-T_n)(1-2 T_n)/\sum_n T_n$. The
functions $A(\omega_1)$ and $B(\omega_2)$ are defined as
\begin{subequations}
\begin{align}
\begin{split}
&A(x_1,x_2) \equiv \int dE f(E) (1-f(E+x_1))(1-f(E+x_2))\\&
\overset{T\rightarrow 0}{\rightarrow}
\theta(x_1)\theta(x_2)\min(x_1,x_2)
\end{split}\\
\begin{split}
&B(x_1,x_2) \equiv \int dE f(E) (1-f(E+x_1))f(E+x_2) \\&\overset{T
\rightarrow 0}{\rightarrow} \theta(x_1) \theta(x_1-x_2)
\min(x_1,x_1-x_2),
\end{split}
\end{align}
\label{eq:ABfunctions}
\end{subequations}
where $f(E)$ is a Fermi function.

In an electric circuit containing reactive elements, the fluctuation
spectra are modified in a frequency dependent way. These types of
modifications can be fairly generally calculated with a Langevin
approach (see for example Refs.~\onlinecite{blanter:1} and
\onlinecite{virtanen:50}; for an exception relevant for the third
cumulant, see Ref.~\onlinecite{beenakker:176802}). In the case of a
time-independent average current, the noise spectra are modified
according to
\begin{equation}
\begin{split}
S_I^{c}(\omega)&=G(\omega)G(-\omega) S_I^{\rm exc}(\omega),\\
\third^c(\omega_1,\omega_2)&=G(\omega_1)G(\omega_2)G(-\omega_1-\omega_2)
\third(\omega_1,\omega_2),
\end{split}
\label{eq:circuiteffect}
\end{equation}
where $G(\omega)$ is a function characterizing the circuit. In what
follows, we choose
\begin{equation}
G(\omega)=1/(1-i\omega/\omega_c), \label{eq:circuitG}
\end{equation}
typical for a circuit with reactive elements next to the point
contact.

Below, we aim to calculate the outcome of these spectra on two
generic detectors coupled to the noise source, a quantum two-level
system and a harmonic oscillator.

\section{Quantum two-level system}
\label{sec:tls}

The general Hamiltonian for a quantum two-level system is
\begin{equation*}
H_{TLS}=-\frac{\tilde{B}_z}{2} \sigma_z - \frac{B_x}{2} \sigma_x,
\end{equation*}
where $\sigma_{z/x}$ are Pauli matrices, and $B_{z/x}$ the effective
magnetic fields. We assume that such a system is coupled to the
noise source via a coupling Hamiltonian
\begin{equation*}
\hcoup=\frac{\hbar}{e}g (I+\delta\hat{I}) \sigma_z.
\end{equation*}
The field pointing to the $y$-direction would not add any more
generality to our model. The average current $I$ can be included in
the classical control field by defining $B_z \equiv
\tilde{B}_z+2\hbar gI/e$, so we can only concentrate on the
fluctuations $\delta \hat{I}$.

The ground and excited states of this system are given by $|0\rangle
= -\beta |\uparrow \rangle + \alpha |\downarrow \rangle$ and $|1
\rangle = \alpha |\uparrow \rangle + \beta |\downarrow \rangle$,
with $\alpha=\cos(\phi/2)$, $\beta=\sin(\phi/2)$ and
$\phi=\arctan(B_z/B_x)$. The energies of these states are
$E_{0/1}=\mp \Omega/2$, $\Omega \equiv \sqrt{B_x^2+B_z^2}$.

Now, the second-order contribution to the excitation rates is
\begin{equation}
\Gamma_{\rm 0 \rightarrow 1}^{(2)}=\frac{g^2}{e^2} |\sigma_z^{01}|^2
S_{\rm noise}(-\Omega/\hbar),
\end{equation}
and the third-order contribution can be obtained from
\begin{equation}
\begin{split}
\Gamma_{\rm 0 \rightarrow 1}^{(3)}= -\frac{g^3}{e^3}
|\sigma_z^{01}|^2 \cos(\phi) \Re\bigg[&\int_{-\infty}^\infty d\omega
\frac{\third(\omega,\Omega/\hbar)}{\omega+\Omega/\hbar-i\eta} \\-&
\int_{-\infty}^\infty d\omega
\frac{\third(\omega,\Omega/\hbar)}{\omega-i\eta}\bigg].
\label{eq:tlsthirdorderrates}
\end{split}
\end{equation}
Here we used the fact that
$\sigma_z^{11}=-\sigma_z^{00}=\cos(\phi)$. Note that the matrix
element $|\sigma_z^{01}|^2=\sin^2(\phi)$ is common for both rates.
The corresponding relaxation rates $\Gamma_{\rm 1 \rightarrow
0}^{(2/3)}$ can be obtained from the excitation rates with the
substitution $\Omega \rightarrow -\Omega$.

In the case of a point contact at a vanishing temperature ($k_B T
\ll eV, \Omega$), the second-order excitation and relaxation rates
are given by
\begin{subequations}
\begin{align}
\Gamma_{\rm 0 \rightarrow 1}^{(2)}&=\frac{g^2}{e^2}|\sigma_z^{01}|^2
F_2 \GN \frac{(e|V|-
\Omega)\theta(e|V|-\Omega)}{1+(\Omega/\hbar \omega_c)^2}\\
\Gamma_{\rm 1 \rightarrow 0}^{(2)}&=\frac{g^2}{e^2}|\sigma_z^{01}|^2
\GN \left[\frac{F_2 (e|V|-\Omega)\theta(e|V|-\Omega) + 2
\Omega}{1+(\Omega/\hbar \omega_c)^2}\right].
\end{align}
\end{subequations}
The calculation of the integrals required for the third-order effect
in this case is detailed in Appendix \ref{app:integrals}. The
general result with an arbitrary ratio between $eV$ and $\hbar
\omega_c$ can be found analytically, but it is too long to be
written down here. In the limit $\Omega < |eV| \ll \hbar \omega_c$
we obtain
\begin{widetext}
\begin{equation}
\Gamma_{\rm 0 \rightarrow 1}^{(3)}= -\frac{g^3}{e^3}
|\sigma_{z}^{01}|^2 \cos(\phi)[I_3-I_1] =\frac{2g^3}{e^2}
|\sigma_{z}^{01}|^2 \cos(\phi) \GN \sgn(V) (F_2-F_3)\left[\Omega
\ln\left(\frac{|eV|-\Omega}{\Omega}\right)+|eV|
\ln\left(\frac{|eV|}{|eV|-\Omega}\right)\right] \label{eq:g3evsmall}
\end{equation}
and $\Gamma_{\rm 1 \rightarrow 0}^{(3)} = - \Gamma_{\rm 0
\rightarrow 1}^{(3)}$. For $|eV| < \Omega$, the contributions to the
excitation rates from the third cumulant vanish, but there is a
contribution to the relaxation rate,
\begin{equation}
\Gamma_{\rm 1 \rightarrow 0}^{(3)}(|eV| < \Omega) = \frac{g^3}{e^2}
|\sigma_z^{01}|^2 \cos(\phi) \GN F_2 \frac{8 \pi \hbar^3 \omega_c^3
\Omega}{4 \hbar^4 \omega_c^4 + 5 \hbar^2 \omega_c^2 \Omega^2 +
\Omega^4}V. \label{eq:g3lowV}
\end{equation}
This contribution becomes small in the limit $\hbar\omega_c \gg
\Omega$.

For $|eV| \gg \Omega$, these rates tend to $\pm \Gamma_{\rm
close}^{(3)}$ with
\begin{equation}
\Gamma_{\rm close}^{(3)}=\frac{2g^3}{e^2} |\sigma_{z}^{01}|^2
\cos(\phi) \GN \sgn(V) (F_2-F_3) \Omega
\left[\ln\left(\frac{|eV|}{\Omega}\right)+1-\frac{\Omega}{2|eV|}\right]+o\left(\left(\frac{\Omega}{|eV|}\right)^2\right).
\label{eq:g3close}
\end{equation}
\end{widetext}
Note that for a tunnel junction, $F_2=F_3$, and there is no
contribution from the third cumulant to the transition rates. For
other types of contacts, the rates are determined according to
$F_2-F_3=2\sum_n T_n^2(1-T_n)/\sum_n T_n$. This is the same as the
zero-frequency Fano factor one would get for an unordered third
cumulant.\cite{saloup06}

In the opposite limit where the voltage by far exceeds the scale
$\hbar \omega_c/e$ set by the circuit, the frequency dependence is
governed by $\omega_c$ and the third-cumulant effect on the
excitation rate is given by
\begin{equation}
\Gamma_{\rm 0 \rightarrow 1}^{(3)} =-\frac{4\pi g^3}{e^2}
|\sigma_{z}^{01}|^2 \cos(\phi) F_3 \frac{\GN \hbar^3\omega_c^3
\Omega}{4 \hbar^4\omega_c^4+5 \hbar^2\omega_c^2 \Omega^2+\Omega^4}V
\label{eq:g3far}
\end{equation}
and again $\Gamma_{\rm 1 \rightarrow 0}^{(3)}=-\Gamma_{\rm 0
\rightarrow 1}^{(3)}$. This is the result one would obtain by
assuming $\third^c(\omega_1,\omega_2) \approx G(\omega_1)
G(\omega_2) G(-\omega_1-\omega_2) \third(0,0)$ as was done in
Ref.~\onlinecite{ojanen:020501}.

\begin{figure}[h]
\centering
\includegraphics[width=\columnwidth]{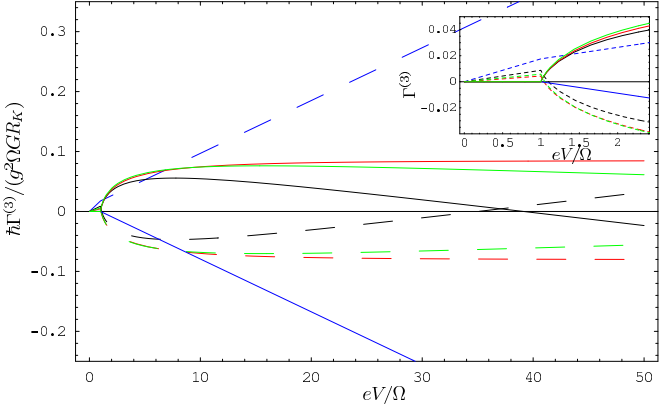}
\caption{(Color online): Third-order contributions to the excitation
(solid lines) and relaxation (dashed lines) rates of a quantum
two-level system coupled to a point contact. The rates are plotted
for four different types of contacts with equal conductance $\GN$:
tunnel contact (blue, $F_2=F_3=1$), dirty interface\cite{schep:3015}
(black, $F_2=1/2$, $F_3=1/4$), diffusive wire\cite{nazarov:134}
(red, $F_2=1/3$, $F_3=1/15$), and a chaotic
cavity\cite{baranger:142} (green, $F_2=1/4$, $F_3=0$). The other
parameters used in this plot are $\phi=\pi/4$ and $\hbar \omega_c=20
\Omega$. Inset shows the low-voltage region. The rates are
proportional to the dimensionless constant $\GN R_K$ where
$R_K=h/e^2$ is the resistance quantum.} \label{fig:thirdorderrates}
\end{figure}

The third-order contributions to the relaxation rates induced by the
noise source are plotted in Fig.~\ref{fig:thirdorderrates} for a few
example cases.

Now let us analyze the detection of noise with the two-level system
via the steady-state occupation probabilities $P_0$ and $P_1=1-P_0$
of the states $|0\rangle$ and $|1\rangle$. These satisfy the
detailed-balance condition, Eq.~\eqref{eq:detailedbalance}.
Analogous to the equilibrium system, we can define an effective
temperature of the quantum two-level system,
\begin{equation}
k_B T_{\rm TLS}=\frac{\Omega}{\ln\left(\frac{P_0}{P_1}\right)} =
\frac{\Omega}{\ln\left(\frac{\Gamma_{1 \rightarrow
0}^{(2)}+\Gamma_{1 \rightarrow 0}^{(3)}}{\Gamma_{0 \rightarrow
1}^{(2)}+\Gamma_{0 \rightarrow 1}^{(3)}}\right)} + o(g^4).
\label{eq:ttls}
\end{equation}
The contributions from the second and third cumulant can be
separated by considering what happens to $T_{\rm TLS}$ when the
voltage across the point contact is reversed. We define the average
temperature $\bar{T}_{\rm TLS} \equiv (T_{\rm TLS}(V)+T_{\rm
TLS}(-V))/2$ and the difference $\Delta T_{\rm TLS} \equiv T_{\rm
TLS}-T_{\rm TLS}(-V)$. In the lowest order in the coupling constant
$g$, the previous is then independent of the third cumulant, whereas
the latter is directly proportional to it.

The limiting case expressions for $\Tav$ and $\dT$ depend on the
relative strengths of the relaxation/excitation rates from the bath
and from the noise source, and on the magnitude of the circuit
frequency scale $\omega_c$. The previous is easiest to characterize
through the ratio of the differences between relaxation and
excitation they cause (this is essentially the "friction" strength
in classical models),\cite{reldirectionnote}
\begin{equation}
\Lambda \equiv \lambda^2 \frac{e^2 (\hbar \omega_c^2 + \Omega^2)
\sin^2(\phi)}{2 \hbar^2 g^2 \GN \omega_c^2 \Omega}.
\end{equation}
For $\Omega \ll k_B T_e,|eV|$, we then get an asymptotic expression
for $\Tav$,
\begin{equation}
k_B \Tav = \frac{2k_B T_e \Lambda + F_2 |eV|}{2(1+\Lambda)} +
\frac{(1-F_2) \Omega}{2(1+\Lambda)} + o\left(\frac{\Omega}{k_B
T_e}\right).
\end{equation}
For $|eV| \gg k_B T_e, \Omega$, there is another asymptotic
expression,
\begin{equation}
k_B \Tav = \frac{F_2 |eV|}{2(1+\Lambda)} + \frac{1-F_2+\Lambda
\coth\left(\frac{\Omega}{2 k_B T_e}\right)}{2(1+\Lambda)}\Omega +
o\left(\frac{\Omega}{|eV|}\right).
\end{equation}
The latter equation is valid for an arbitrary ratio between $\Omega$
and $k T_e$.

\begin{figure}[h]
\centering
\includegraphics[width=\columnwidth]{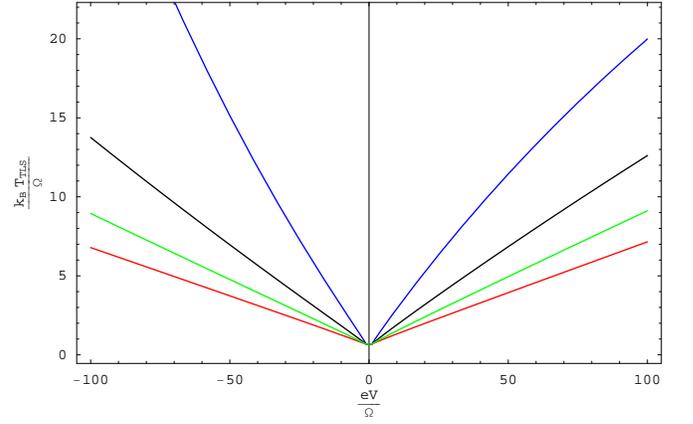}
\caption{Effective temperature of the two-level system as a function
of the voltage through different types of point contacts coupled to
it. The different curves correspond, from top to bottom, to a tunnel
contact (blue), dirty interface (black), diffusive wire (green) and
a chaotic cavity (red). The behavior is mostly dictated by the Fano
factor $F_2$. Other parameters in the plot are $g=0.05$, $\hbar
\omega_c=20 \Omega$, $\phi=\pi/4$, $\Lambda=1$, $k_B T_e=\Omega$,
and $\GN=1/R_K$.} \label{fig:Ttlsvsfano}
\end{figure}

\begin{figure}[h]
\centering
\includegraphics[width=\columnwidth]{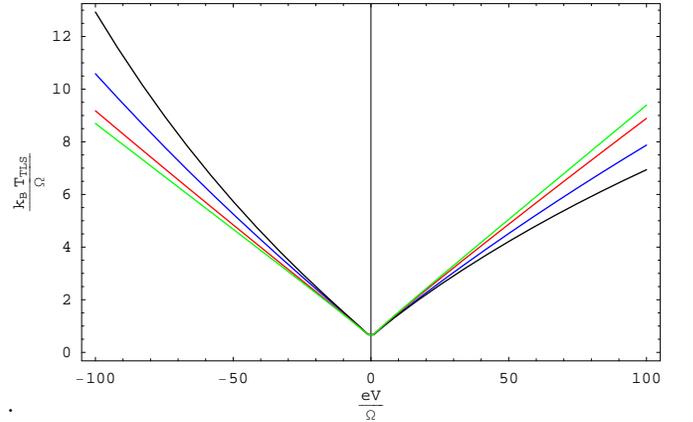}
\caption{Effective temperature of the two-level system as a function
of the voltage through a diffusive point contact coupled to it via
circuits characterized with different $\omega_c$. In the right, from
top to bottom: $\hbar\omega_c/\Omega=100$ (green), 10 (red), 0.1
(blue) and 1 (black). Other parameters are as in
Fig.~\ref{fig:Ttlsvsfano}.} \label{fig:Ttlsvsomegac}
\end{figure}

The effective temperature $T_{\rm TLS}$ as a function of the voltage
$V$ is plotted in Fig.~\ref{fig:Ttlsvsfano} for a few types of
junctions and in Fig.~\ref{fig:Ttlsvsomegac} for a few values of
$\omega_c$.

\begin{figure}[h]
\centering
\includegraphics[width=\columnwidth]{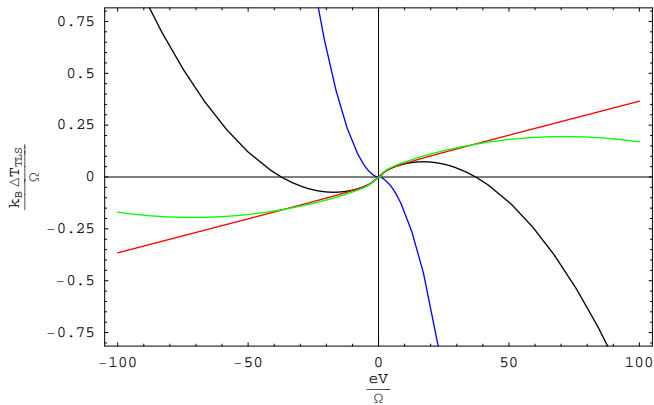}
\caption{Asymmetric part of the temperature $\dT$ as a function of
the voltage over the point contact for different types of point
contacts. In the right, from top to bottom: chaotic cavity (red),
diffusive wire (green), dirty interface (black) and a tunnel
junction (blue). Other parameters are as in
Fig.~\ref{fig:Ttlsvsfano}.} \label{fig:dTvsfano}
\end{figure}

\begin{figure}[h]
\centering
\includegraphics[width=\columnwidth]{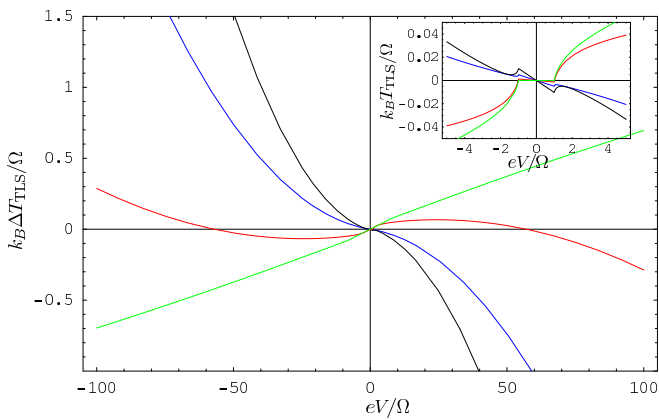}
\caption{Asymmetric part of the temperature $\dT$ as a function of
the voltage over a diffusive point contact with different
frequencies characterizing the circuit: in the right, from top to
bottom: $\hbar \omega_c/\Omega=100$ (green), 10 (red), 0.1 (blue),
and 1 (black). Other parameters are as in Fig.~\ref{fig:Ttlsvsfano}.
The curve for $\hbar \omega_c=100\Omega$ is in the logarithmic
regime, Eq.~\eqref{eq:dTclose}, for all plotted voltages, and it is
characterized by the Fano factor $F_2-F_3$. The curves for $\hbar
\omega_c \le 1$ are in the linear regime, Eq.~\eqref{eq:dTfar}, and
the case $\hbar \omega_c=10 \Omega$ has a crossover between the two.
Inset shows the low-voltage regime, reflecting the behavior
described in Eq.~\eqref{eq:g3lowV}.} \label{fig:dTdiffvsomegac}
\end{figure}

The antisymmetric part $\dT$ of the temperature with respect to the
voltage through the scatterer depends strongly on whether the
frequency dependence is governed by the voltage or by the circuit.
In the previous case, for $\Omega \ll k_B T_e, |eV| \ll \hbar
\omega_c$ (noise source "close" to the detector), the asymptotic
expression for $\dT$ is
\begin{equation}
\begin{split}
&k_B \dT =
\sgn(V)\frac{2g(F_2-F_3)\cos(\phi)}{(1+\Lambda)^2}\\&\times\left(F_2
|eV|+2 k_B T_e
\Lambda\right)\left[\ln\left(\frac{|eV|}{\Omega}\right)+1\right] +
o\left(\frac{\Omega}{|eV|}\right). \label{eq:dTclose}
\end{split}
\end{equation}
For a noise source placed "far" from the detector, i.e., $\Omega,
\hbar \omega_c \ll |eV|, k T_e$, we get
\begin{equation}
\begin{split}
&k_B \dT = -V \cos(\phi)\frac{4 gF_3 \pi \hbar \omega_c}{(4 \hbar^2
\omega_c^2 + \Omega^2) (1+\Lambda)^2}\\&\times \left(F_2 |eV|+2 k_B
T_e \Lambda\right) + o\left(\frac{\Omega,\hbar
\omega_c}{|eV|}\right).\label{eq:dTfar}
\end{split}
\end{equation}
These limits are illustrated in Figs.~\ref{fig:dTvsfano} and
\ref{fig:dTdiffvsomegac} which show the temperature difference $\dT$
as a function of the voltage over the point contact for different
types of point contacts and for different $\omega_c$, respectively.
For a tunnel junction, $\dT/V$ is negative for all values of the
voltage as the logarithmic term in $V$ (Eq.~\eqref{eq:dTclose}) is
absent, whereas for a chaotic cavity $F_3=0$ and the absence of the
linear term (Eq.~\eqref{eq:dTfar}) leads to a positive definite
$\dT/V$. For other types of junctions, there is a crossover from
positive $\dT/V$ to a negative $\dT/V$ as $eV$ roughly crosses
$\hbar \omega_c$.

The relative temperature change $\dT/\Tav$ is thus logarithmic in
$V$ for $\Omega, k_B T_e \ll |eV| \ll \hbar \omega_c$ and linear for
$|eV| \gg \Omega, k_B T_e, \hbar \omega_c$.

\section{Harmonic oscillator}
\label{sec:ho}

Below, we discuss two schemes for using a quantum harmonic
oscillator as a detector of the current fluctuations. The first
scheme couples a displaced oscillator to the fluctuations through
the position operator $\hat{x}$, and the second scheme through the
second power of the momentum operator, $\hat{p}^2$. It turns out
that at least up to the third order in the coupling operators $g_i$,
$\lambda_i$, the first scheme is insensitive to the third cumulant,
whereas the latter follows quite closely the qubit scheme. First, we
note that a simple harmonic oscillator coupled to the fluctuations
via its position operator, i.e., $\hat{A}=\hat{x}$ directly implies
$\Gamma^{(3)}=0$ as Eq.~\eqref{eq:generalthirdorder} requires that
all the matrix elements between the initial, the final, and one or
more intermediate states are nonzero. This cannot be satisfied as
$\hat{x}$ couples only the neighboring states.

\subsection{Displaced harmonic oscillator coupled to the position
operator}

This restriction may be overcome by considering a {\it displaced}
harmonic oscillator, so that the effective coupling is to
$\hat{x}-x_0$, where $x_0$ is a scalar displacement. However, one
can quite generally show that even in this case the resulting
$\Gamma^{(3)}=0$.

To be specific, consider a harmonic oscillator coupled to the
fluctuating current. The Hamiltonian for the oscillator is
\begin{equation*}
H_{\rm HO}=-\frac{\hbar}{2m}\partial_{x'}^2 + \frac{1}{2}m
\omega_0^2 \hat{x}'^2.
\end{equation*}
Assume the fluctuating current is coupled to the
$\hat{x}'$-coordinate, i.e., the coupling is described by
\begin{equation*}
H_{\rm coup} = \frac{g}{e} (I_b+\delta \hat{I})
\sqrt{2m\hbar\omega_0}\hat{x}'.
\end{equation*}
Due to the average current term $I_b$, the oscillator potential
minimum is shifted from $x=0$ to $x_0=-\sqrt{2\hbar} g I_b/(e
\sqrt{m\omega_0^3})$. Defining a new displaced operator $\hat{x}
\equiv \hat{x}'-x_0$ we get, neglecting the unimportant scalar
terms,
\begin{equation}
\begin{split}
H_{\rm HO}+H_{\rm coup} =& -\frac{\hbar}{2m} \partial_x^2 +
\frac{1}{2} m \omega_0^2 \hat{x}^2 \\&+ g
\frac{\sqrt{2m\omega_0}}{e}\delta \hat{I} \left(\hat{x}+x_0\right).
\end{split}
\end{equation}
In what follows, we assume that $I_b$ can be controlled separately
from the current flowing through the noise source. This type of a
separation of average and noise currents was discussed for example
in Ref.~\onlinecite{pekola:197004}.

We proceed in the usual way by defining the harmonic oscillator
annihilation and creation operators,
\begin{equation*}
\begin{cases}
\hat{a}\\
\hat{a}^\dagger
\end{cases}
= \frac{1}{\sqrt{2 \hbar m \omega_0}}\left(m\omega_0 \hat{x} \mp
\hbar \partial_x\right).
\end{equation*}
Now the energy eigenstates $|n\rangle$ of the "average" Hamiltonian
are those of the number operator, $\hat{N} |n\rangle =
\hat{a}^\dagger \hat{a} |n\rangle = |n \rangle$. The fluctuations
are coupled to the operator $\hat{A} \equiv \sqrt{2m\hbar
\omega_0}(\hat{x}+x_0)$. This has a finite matrix element $A_{n,n\pm
1}$ between the neighboring states due to the operator $\hat{x}$.
Moreover, the displacement $x_0$ makes the diagonal matrix element
$\hat{A}_{nn}$ also finite, and independent of the level index $n$.
Similar to the two-level system, we can write the second-order
excitation and relaxation rates due to the external noise from a
point contact at $k_B T \ll |eV|, \hbar \omega_0$,
\begin{subequations}
\begin{align}
\begin{split}
&\Gamma_{n \rightarrow n+1}^{(2)} = \frac{g^2}{e^2}(n+1)
S_{I}(-\omega_0)\\&=\frac{g^2}{e^2}(n+1) \GN F_2 (|eV|-\hbar
\omega_0)\theta(|eV|-\hbar \omega_0)
\end{split}\\
\begin{split}
&\Gamma_{n+1 \rightarrow n}^{(2)} = \frac{g^2}{e^2}(n+1)
S_{I}(\omega_0) \\&= \frac{g^2}{e^2}(n+1) \GN [F_2(|eV|-\hbar
\omega_0)\theta(|eV|-\hbar \omega_0) + 2 \hbar \omega_0].
\end{split}
\end{align}
\end{subequations}
The third-order contribution to the excitation rate is
\begin{equation}
\begin{split}
\Gamma_{n \rightarrow n+1}^{(3)} = \frac{2g^4 I_b}{e^4 \omega_0}
(n+1) \Re\bigg[&\int_{-\infty}^\infty d\omega
\frac{\third(\omega,\Omega/\hbar)}{\omega+\Omega/\hbar-i\eta} \\+
&\int_{-\infty}^\infty d\omega
\frac{\third(\omega,\Omega/\hbar)}{\omega-i\eta}\bigg]
\end{split}
\label{eq:hothirdorder}
\end{equation}
and the relaxation rate $\Gamma_{n+1 \rightarrow n}^{(3)}$ can be
obtained by replacing $\omega_0$ by $-\omega_0$ inside the integrals
(but not in the prefactor).

These integrals are the same as for the quantum two-level system,
Eq.~\eqref{eq:tlsthirdorderrates}, up to a sign between them. But as
shown in Appendix \ref{app:integrals}, the two integrals give
exactly the opposite contribution under quite general conditions,
and therefore $\Gamma^{(3)}$ vanishes.

\subsection{Coupling to the square of the momentum or position
operator} \label{subs:hosquare}

Assume one could vary the mass of the harmonic oscillator via the
fluctuating current. In this case, the coupling to the fluctuations
would be of the form\cite{secondterm}
\begin{equation}
H_{\rm coup} = \frac{\delta \hat{m}(\delta \hat{I})}{2m^2} \hat{p}^2
= -g \delta \hat{I} (\hat{a}^\dagger-\hat{a})^2.
\end{equation}
Such an operator $\hat{A}=(\hat{a}^\dagger-\hat{a})^2$ has finite
matrix elements between next-nearest neighbor energy levels $n$ and
$n+2$ of the oscillator, $A_{n,n+2}=\sqrt{(n+1)(n+2)}$ and diagonal
matrix elements, $A_{n,n}=-2n-1$. To be able to further describe the
system with an effective temperature, we assume that the coupling to
the bath is much weaker than the coupling to the noise source, and
the previous can hence be neglected. Now the second-order
contribution to the transition rates due to the current fluctuations
at $T=0$ are
\begin{subequations}
\begin{align}
\begin{split}
&\Gamma_{n \rightarrow n+2}^{(2)} = \frac{g^2}{e^2}(n+1)(n+2)
S_{I}(-2\omega_0)\\&=\frac{g^2}{e^2}(n+1)(n+2) \GN F_2 (|eV|-2\hbar
\omega_0)\theta(|eV|-2\hbar \omega_0)
\end{split}\\
\begin{split}
&\Gamma_{n+2 \rightarrow n}^{(2)} = \frac{g^2}{e^2}(n+1)(n+2)
S_{I}(2\omega_0) \\&= \frac{g^2}{e^2}(n+1)(n+2) \GN [F_2(|eV|-2\hbar
\omega_0)\theta(|eV|-2\hbar \omega_0) + 4 \hbar \omega_0].
\end{split}
\end{align}
\end{subequations}
The third-order contribution to the excitation rate is
\begin{equation*}
\begin{split}
&\Gamma_{n \rightarrow n+2}^{(3)} = -\frac{g^3}{e^3} (n+1)(n+2)
\Re\bigg[(2n+1)\\&\times\int_{-\infty}^\infty d\omega
\frac{\third(\omega,2\omega_0/\hbar)}{\omega+2\omega_0-i\eta} +
(2n+3)\int_{-\infty}^\infty d\omega
\frac{\third(\omega,2\omega_0/\hbar)}{\omega-i\eta}\bigg] \\&\equiv
-\frac{g^3}{e^3}(n+1)(n+2)\left[(2n+1)I_3+(2n+3)I_1\right]
\end{split}
\end{equation*}
and the relaxation rate $\Gamma_{n+2 \rightarrow n}^{(3)}$ can be
obtained by replacing $\omega_0$ by $-\omega_0$. Using the fact that
$I_3=-I_1$ (see Appendix \ref{app:integrals}), we get
\begin{equation}
\Gamma_{n \rightarrow n+2}^{(3)} = -\frac{2g^3}{e^3} (n+1)(n+2)
I_1(2\omega_0).
\end{equation}
The third-order rate has thus the same level-dependent prefactor as
the second-order rate. Therefore, we can again define an effective
temperature, which now is of the form
\begin{equation}
k_B T_{\rm ho}=\frac{2\hbar\omega_0}{\ln\left(\frac{\Gamma_{n+2
\rightarrow n}^{(2)}+\Gamma_{n+2 \rightarrow n}^{(3)}}{\Gamma_{n
\rightarrow n+2}^{(2)}+\Gamma_{n \rightarrow n+2}^{(3)}}\right)}.
\end{equation}
This is independent of the level index $n$, provided the
fluctuations coupling linearly to $\hat{x}$ or $\hat{p}$ can be
neglected.

Because of the similar form of the rate expressions as in the qubit
case, the behavior of the effective temperature is similar to the
qubit, provided the prefactor $\cos(\phi)$ in
Eqs.~(\ref{eq:g3close},\ref{eq:g3far},\ref{eq:dTclose},\ref{eq:dTfar})
is replaced by $-1$ and the matrix element $\sigma_z^{01}$ by
$(n+1)(n+2)$. In the case of coupling to $\hat{x}^2$ instead of
$\hat{p}^2$ (i.e., varying the spring constant rather than the
mass), the only difference is the inverted sign of the
third-cumulant contributions to the rates.

\section{Experimental detector realizations}

\begin{figure}[h]
\centering
\includegraphics[width=\columnwidth]{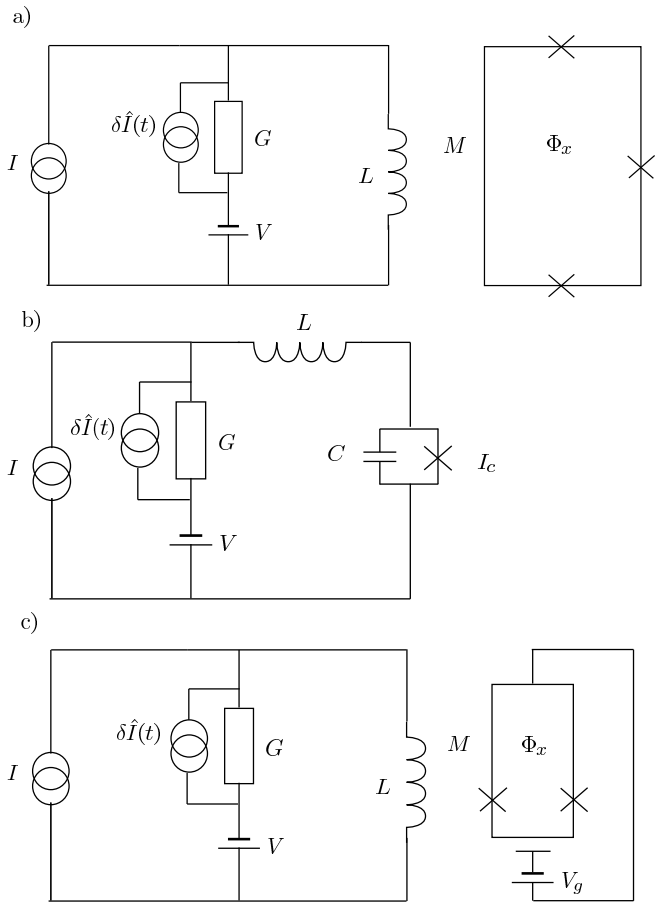}
\caption{Fluctuation measurement schemes using superconducting
qubits: a) Persistent current qubit, b) phase qubit, and c) charge
qubit.} \label{fig:qubitschemes}
\end{figure}

To exemplify the measurement of the third cumulant through the
polarization of a qubit coupled to the fluctuating current, we
consider three specific examples: a persistent current (or a flux)
qubit,\cite{majer:090501,chiorescu:1869} a phase
qubit,\cite{martinis:094510,martinis:117901} and a charge
qubit.\cite{nakamura:786} The fluctuation measurement schemes with
these qubits are illustrated in Fig.~\ref{fig:qubitschemes}. Apart
from the measurement schemes, we aim to discuss typical values for
the coupling constant $g$, level splitting $\Omega$, circuit
frequency scale $\omega_c$, and the dimensionless constant $\Lambda$
characterizing the ratio between the intrinsic and induced noise.

In all cases, the level splitting $\Omega/\hbar$ turns out to be of
the order of a few GHz at minimum. Also, for all systems the
relevant reactive element causing dispersion of noise is the
inductance $L$ of the lines feeding the noise current, and therefore
the frequency scale $\omega_c=1/(\GN L)$, where $\GN$ is the
conductance of the shot noise source. As $L$ is typically of the
order of some 10 pH, and $R$ may vary between some 10 $\Omega$ to
some 100 k$\Omega$, we have $\omega_c$ ranging between 10$^3$ to
10$^6$ GHz. This means that unless special care is taken to make a
large inductance, $\hbar\omega_c/\Omega$ is at least a few hundred.

However, note that at least in diffusive wires and chaotic cavities,
there is an additional frequency scale given by the inverse dwell
time $1/\tau_D$ or the inverse screening time
$1/\tau_{sc}$.\cite{nagaev:176804,hekking:056603,pilgram:045304} For
example, for a diffusive wire of length $L=1 \mu$m and diffusion
constant $D=100$ cm$^2$/s, we have $1/\tau_D \approx 10$ GHz. When
$1/\tau_D$ or $1/\tau_{sc}$ is less than
$\min(|eV|/\hbar,\omega_c)$, these have to be taken into account
separately.

With the qubits, the detection takes place by applying a steady
current through the noise source, and measuring the occupation
number $\rho_{11}$ of the higher qubit state many times, i.e.,
averaging over the realizations of current fluctuations. The
effective temperature is then
\begin{equation*}
k_B T_{\rm eff} = \Omega
\ln\left(\frac{1-\rho_{11}}{\rho_{11}}\right).
\end{equation*}
The third-cumulant effect can be controlled by tuning the phase
$\phi$ through the average fields $B_x$ and $B_z$. When either of
these fields vanishes, also the third-cumulant effect should vanish.

The temperature measurement of the harmonic oscillator depends on
its realization: For a true oscillator based on a resonant
$LC$-circuit, the temperature should be measured either by measuring
the current noise power in the oscillator, or coupling it to a
nonlinear system, say a SQUID, and measuring its response. If the
realization is a current-biased Josephson junction, the temperature
detection can be done via the measurement of the thermal escape rate
as in Ref.~\onlinecite{pekola:197004}.

Voltage fluctuations could also be coupled to the qubits or
oscillators. Typically these would couple to the perpendicular
external field component compared to the current fluctuations.
However, the frequency dependence of the third cumulant of voltage
fluctuations is not known, and the response of the qubits might in
this case be somewhat different.

\subsection{Persistent current qubit}

In the persistent current qubit, a fluctuating current is easiest to
couple to the qubit current through the mutual inductance $M$ as in
Fig.~\ref{fig:qubitschemes}a.\cite{majer:090501} In the basis
defined by the clockwise and anticlockwise current, this corresponds
to coupling to $\sigma_z$. In this case, the coupling constant $g$
is of the order of
\begin{equation*}
g_{pcq} = \frac{e}{\hbar} M \Delta I_{qb} \sim \frac{e}{\hbar} M
I_c,
\end{equation*}
where $\Delta I_{qb}$ is the different between the currents
corresponding to the two qubit states. This is of the order of the
critical current $I_c$ of the Josephson junctions. Using $M \approx
2$ pH and $I_c \approx 3$ $\mu$A close to the experimental
values,\cite{majer:090501,chiorescu:1869} we get $g \approx 0.01$.
Finally, with the intrinsic relaxation time $T_1 = 1$ ms,
$\Omega/\hbar =$10 GHz, $R=R_K$ and $\phi=0$, we get $\Lambda
\approx 0.003$, i.e., the intrinsic bath effect is almost
negligible.

\subsection{Phase qubit}

With the scheme depicted in Fig.~\ref{fig:qubitschemes}b, the
external current fluctuations can be again coupled to the diagonal
element of the qubit Hamiltonian, i.e., $\sigma_z$. The coupling
strength is\cite{martinis:094510}
\begin{equation*}
g_{pq}= \frac{e}{\hbar}\frac{\partial E_{10}}{\partial I_b},
\end{equation*}
where $E_{10}$ is the level separation between the qubit states. For
the bias current close to $I_c$, this is given by
\begin{equation*}
E_{10} \approx \hbar \omega_p(I) \left(1-\frac{5}{36}\frac{\hbar
\omega_p(I)}{\Delta U(I)}\right),
\end{equation*}
where $\omega_p(I)=2^{1/4} \sqrt{8 E_J E_c} (1-I/I_c)^{1/4}$ and
$\Delta U(I)=4\sqrt{2}/3 E_J (1-I/I_c)^{3/2}$, $E_J=\hbar I_c/(2e)$
and $E_C=e^2/(2C)$ are the Josephson and charging energies of the
junction, and $I_c$ is its critical current. Thus we have
\begin{equation*}
g_{pq} = - \frac{5 \tilde{e}_C + 3 \cdot 2^{3/4}
\sqrt{\tilde{e}_C}(1-i_b)^{5/4}}{6 (1-i_b)^2}
\end{equation*}
where $i_b=I/I_c$, and $\tilde{e}_c=E_C/E_J$. With the
values\cite{martinis:117901} $C=6$ pF and $I_c=21$ $\mu$A and $i_b
\approx 0.99$, we get $g_{pq} \approx 0.1$. In such qubits, the
relaxation due to the intrinsic bath is slower than $\omega_p/1000$,
which implies $\Lambda < 0.1$. Therefore, the intrinsic bath should
also here be negligible.

\subsection{Charge qubit}

Coupling a charge qubit in a form of the Cooper pair
box\cite{nakamura:786} to current fluctuations takes place via the
operator $\sigma_x$ in the natural charge basis of the qubit. This
is accomplished by coupling the fluctuations to the flux controlling
the Josephson energy of the Cooper pair box. Due to this slightly
different type of a coupling, the phase $\phi$ should be defined as
$\phi=\arctan(B_x/B_z)$, but otherwise the rate expressions stay the
same. In this case, the coupling strength is given
by\cite{makhlin:357}
\begin{equation*}
g=\frac{e}{\hbar} M \frac{d I_c(\Phi_x)}{d\Phi_x},
\end{equation*}
where $M$ is the mutual inductance and $I_c(\Phi_x)$ is the critical
current of the box at the average external flux $\Phi_x$. With $M=2$
pH and $I_c=300$ nA, $g$ can be made to vary between almost zero
(the "sweet spot" where the first derivative of $I_c(\Phi_x)$ vs.
$\Phi_x$ vanishes) to some 10$^{-3}$. A higher coupling strength can
be obtained by increasing $I_c$ and thereby going away from the
strict charge basis as in Ref.~\onlinecite{vion:886}.

\begin{figure}[h]
\centering \includegraphics[width=\columnwidth]{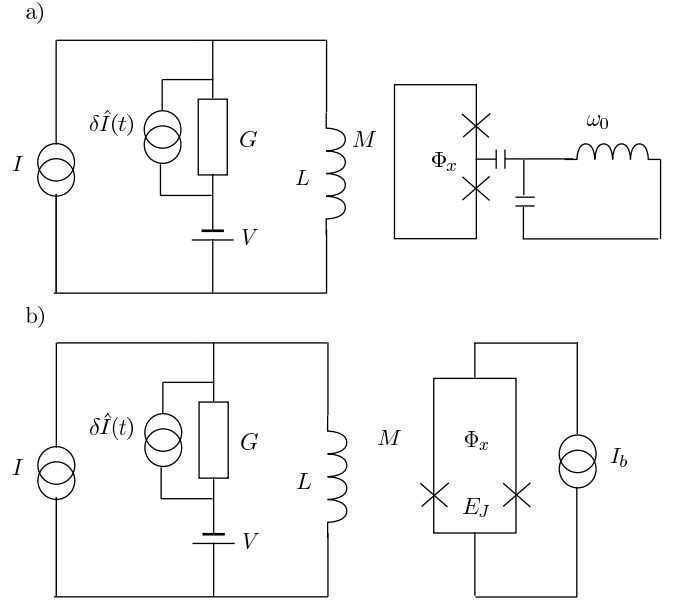}
\caption{Fluctuation measurements with oscillators whose mass or
spring constant is driven with the fluctuations: a) Cooper-pair box
and b) driven Josephson junction. In both cases, the Josephson
energy is tuned with the fluctuations. In a), the difference to a
charge qubit is in the fact that the transistor is placed as a part
of a resonant circuit.} \label{fig:oscillators}
\end{figure}

\subsection{Cooper pair box in a resonant circuit}
\label{subs:cpbox}

Coupling the fluctuations to a flux controlling the Josephson
coupling of a Cooper pair box placed in a resonant circuit, one may
control the effective capacitance of the
resonator.\cite{roschier:024530,sillanpaa:206806,duty:206807} Such a
setup corresponds to that studied in Subs.~\ref{subs:hosquare}. The
coupling strength $g$ depends on the relative ratio between the
Josephson and charging energies, $E_J$ and $E_C$ of the Cooper pair
box. For $E_J=10 E_C$, the relative capacitance modulation $\Delta
C/C$ can be of the order of 0.05 for a change $\Delta
\Phi_x=\hbar/2e$ in the external flux.\cite{sillanpaa:206806} The
coupling constant in this case is
\begin{equation*}
g=-\frac{\Delta C}{C} \frac{e^2 M \omega_0}{2\hbar}.
\end{equation*}
For $M=100$ pH and $\omega_0=$100 GHz, we would hence obtain $g \sim
10^{-4}$. This is quite a small value, and using this scheme would
require optimization of the $E_J/E_C$-ratio and maximizing both $M$
and $\omega_0$.

\subsection{Current biased SQUID}

Instead of directly controlling the bias current of the Josephson
junction as in the phase qubit scheme above, one may also envisage
coupling the fluctuations inductively to control the flux in a SQUID
as in Fig.~\ref{fig:oscillators}b. The Hamiltonian of a biased
symmetric SQUID is
\begin{equation*}
H_{JJ}=-4E_C \partial_\varphi^2 -E_J
\cos\left(2\pi\frac{\Phi_x}{\Phi_0}\right)\cos(\varphi)-\frac{\hbar}{2e}I_b\varphi,
\end{equation*}
where $\varphi$ is the phase across the SQUID, $I_b$ is the bias
current, $\Phi_x$ is the external flux through the loop, and
$\Phi_0=h/(2e)$ is the flux quantum. We assume the self-inductance
of the loop small enough, so that it can be neglected. For $I_b$
much lower than the critical current $I_C(\Phi_x)=2e/\hbar E_J
\cos(2\pi \Phi_x/\Phi_0)$ of the SQUID, and $E_C$, $k_B T_{JJ} \ll
E_J$, we can neglect the driving term and expand the term
$\cos(\varphi)\approx 1-\varphi^2/2$. As a result, we get a harmonic
oscillator Hamiltonian with the mass given by the capacitance $C$,
and the spring constant given by the Josephson inductance
$L_J=\hbar/(2e) I_C$. Now connecting the current fluctuations to the
flux $\Phi_x=M(I+\delta I(t))$, we can vary the inductance term,
i.e., couple to $\hat{x}^2$ of the harmonic oscillator. The coupling
constant is given by\cite{jjsecondterm}
\begin{equation*}
g_{JJ}=-\frac{Me^2\omega_p}{2\hbar}
\sin\left(2\pi\frac{MI}{\Phi_0}\right), \label{eq:jjcp}
\end{equation*}
where $\omega_p=\sqrt{8E_JE_C}/\hbar$ is the plasma frequency of the
SQUID at $\Phi_x=0$. With $M=20$ pH, $\omega_p=$200 GHz, we get
$g=10^{-3}$. Hence, the current biased SQUID can be used for the
detection of the third cumulant in the harmonic mode, but the
parameters $M$ and $\omega_p$ need to be optimized to quite high
values in order to obtain large enough coupling strength.

Another way to use the SQUID in the harmonic mode would be to place
it in a resonant circuit, and use the modulation of the Josephson
inductance for the fluctuation measurement. In this case, the scheme
would be similar to that presented in Subs.~\ref{subs:cpbox}, but
the coupling would be to $\hat{x}^2$ rather than $\hat{p}^2$. The
coupling constant would be again given by Eq.~\eqref{eq:jjcp}, but
now $\omega_p$ should be replaced with the resonance frequency of
the circuit.

\section{Discussion}
We suggest to use the excitation and relaxation in quantum two-level
systems or harmonic oscillators for measuring the third cumulant of
current fluctuations in a short contact. When coupled to the driven
non-Gaussian fluctuations, the static density matrix of these
systems reveals information on the frequency dependence of these
fluctuations. The third cumulant can be read from the change in the
effective temperature of these systems upon reversing the polarity
of the bias across the contact. As the measured signal is inherently
quantum, the ordering of the current operators turns out to be
important. Depending on the relative ratio of the frequency scales
of the system, given by the voltage, $eV/\hbar$, and the circuit,
$\omega_c$, the measured Fano factor for the third cumulant is
either $\sum_n T_n(1-T_n)(1-2T_n)/\sum_n T_n$ ($eV \gg \hbar
\omega_c$, Eqs.~\eqref{eq:g3evsmall} and \eqref{eq:dTclose}) or
$2\sum_n T_n^2 (1-T_n)/\sum_n T_n$ ($eV \ll \hbar \omega_c$,
Eqs.~\eqref{eq:g3far} and \eqref{eq:dTfar}). This slightly resembles
the reasoning in Ref.~\onlinecite{lesovik:393}, where the measured
Fano factor depends on "how far" the detector is placed from the
current path.

\section*{Acknowledgements}
We thank Joachim Ankerhold, Pertti Hakonen, Matthias Meschke, Jukka
Pekola and Peter Samuelsson for enlightening discussions. This work
was supported by the Academy of Finland.

\appendix

\section{Integrals required for the third-order rates}
\label{app:integrals}

\begin{widetext}
The third-order contributions to the rates depend on the integrals
of the form
\begin{equation*}
\begin{split}
I_{1/4} &\equiv \Re\left[\int_{-\infty}^\infty
\frac{\third(\omega,\pm \omega_0)}{\omega} d\omega\right] =
\Re\left[\int_0^\infty \frac{\third(\omega,\pm
\omega_0)-\third(-\omega,\pm \omega_0)}{\omega} d\omega\right]\\
I_{2/3} &\equiv \Re\left[\int_{-\infty}^\infty
\frac{\third(\omega,\mp \omega_0)}{\omega\mp \omega_0}
d\omega\right]=\Re\left[\int_{0}^\infty \frac{\third(\omega \mp
\omega_0,\mp \omega_0)-\third(-\omega \mp \omega_0,\mp
\omega_0)}{\omega} d\omega\right].
\end{split}
\end{equation*}
\end{widetext}
Quite generally, the frequency dependent third cumulant considered
in this manuscript satisfies
$\third(\omega_1,\omega_2)=\third(-\omega_1-\omega_2,\omega_2)$.
This is valid for a point contact at arbitrary temperature
(Eqs.~\eqref{eq:pcthird} -- \eqref{eq:ABfunctions}) in the presence
($\third^c$ from Eq.~\eqref{eq:circuiteffect}) or absence of the
external circuit. Using this symmetry, it is straigthforward to show
that $I_2=-I_4$ and $I_3=-I_1$.

The integrals can be calculated at $T=0$ by evaluating
$(\third(\omega,\pm \omega_0)-\third(-\omega,\pm \omega_0))/\omega$
by parts, $\omega \in [0,\min(\omega_0,|eV|/\hbar-\omega_0)]$,
$\omega \in
[\min(\omega_0,|eV|/\hbar-\omega_0),\max(\omega_0,|eV|/\hbar-\omega_0)]$,
$\omega \in [\max(\omega_0,|eV|/\hbar-\omega_0),|eV|/\hbar]$ and
finally $\omega > |eV|/\hbar$. In the presence of the external
circuit, described with Eq.~\eqref{eq:circuitG}, the resulting
analytic expressions for $I_i$ are very long, even at zero
temperature. However, they have rather simple limits depending on
the relation between $|eV|$ and $\hbar \omega_0$, i.e., which of
these scales gives the cutoff for the integrals. For all $\hbar
\omega_c$, the integrals are finite only for $|eV|
> \Omega$. Moreover, in both limits (but not generally), $I_1=I_2$.
For $|eV| \ll \hbar \omega_c$ we get for $T=0$
\begin{equation}
\begin{split}
I_1=e \GN (F_2-F_3) \sgn(V) \bigg[&\Omega \ln\left(\frac{|eV|-\hbar
\omega_0}{\omega_0}\right)\\+& |eV| \ln\left(\frac{|eV|}{|eV|-\hbar
\omega_0}\right)\bigg].
\end{split}
\end{equation}
For $|eV| \gg \hbar \omega_c$, the integral can be evaluated by
assuming a frequency independent intrinsic third cumulant,
$\third(\omega_1,\omega_2)=G(\omega_1)G(\omega_2)G(-\omega_1-\omega_2)
F_3 e \GN V$ as was done in Ref.~\onlinecite{ojanen:020501}. This
yields
\begin{equation}
I_1=-e \GN F_3\frac{2\pi \omega_c^3 \omega_0}{4 \omega_c^4 + 5
\omega_c^2 \omega_0 + \omega_0^2}.
\end{equation}
This result can be found fairly straigthforwardly via the residue
theorem.


\begin{thebibliography}{47}
\expandafter\ifx\csname
natexlab\endcsname\relax\def\natexlab#1{#1}\fi
\expandafter\ifx\csname bibnamefont\endcsname\relax
  \def\bibnamefont#1{#1}\fi
\expandafter\ifx\csname bibfnamefont\endcsname\relax
  \def\bibfnamefont#1{#1}\fi
\expandafter\ifx\csname citenamefont\endcsname\relax
  \def\citenamefont#1{#1}\fi
\expandafter\ifx\csname url\endcsname\relax
  \def\url#1{\texttt{#1}}\fi
\expandafter\ifx\csname urlprefix\endcsname\relax\def\urlprefix{URL
}\fi \providecommand{\bibinfo}[2]{#2}
\providecommand{\eprint}[2][]{\url{#2}}

\bibitem[{\citenamefont{Nazarov}(2003)}]{nazarovbook}
\bibinfo{editor}{\bibfnamefont{Y.}~\bibnamefont{Nazarov}}, ed.,
  \emph{\bibinfo{title}{Quantum Noise in Mesoscopic Physics}},
  vol.~\bibinfo{volume}{97} of \emph{\bibinfo{series}{Nato Science Series II}}
  (\bibinfo{publisher}{Kluwer}, \bibinfo{address}{Dodrecht},
  \bibinfo{year}{2003}).

\bibitem[{\citenamefont{Belzig}(2003)}]{belzig03}
\bibinfo{author}{\bibfnamefont{W.}~\bibnamefont{Belzig}}, in
  \emph{\bibinfo{booktitle}{Proceedings of Summer School/Conference on
  Functional Nanostructures}} (\bibinfo{year}{2003}),
  \bibinfo{note}{[cond-mat/0312180]}.

\bibitem[{\citenamefont{Levitov et~al.}(1996)\citenamefont{Levitov, Lee, and
  Lesovik}}]{levitov:4845}
\bibinfo{author}{\bibfnamefont{L.~S.} \bibnamefont{Levitov}},
  \bibinfo{author}{\bibfnamefont{H.}~\bibnamefont{Lee}}, \bibnamefont{and}
  \bibinfo{author}{\bibfnamefont{G.~B.} \bibnamefont{Lesovik}},
  \bibinfo{journal}{J. Math. Phys.} \textbf{\bibinfo{volume}{37}},
  \bibinfo{pages}{4845} (\bibinfo{year}{1996}).

\bibitem[{\citenamefont{Ojanen and Heikkil\"a}(2006)}]{ojanen:020501}
\bibinfo{author}{\bibfnamefont{T.}~\bibnamefont{Ojanen}} \bibnamefont{and}
  \bibinfo{author}{\bibfnamefont{T.~T.} \bibnamefont{Heikkil\"a}},
  \bibinfo{journal}{Phys. Rev. B} \textbf{\bibinfo{volume}{73}},
  \bibinfo{eid}{020501(R)} (\bibinfo{year}{2006}).

\bibitem[{\citenamefont{Salo et~al.}(2006)\citenamefont{Salo, Hekking, and
  Pekola}}]{saloup06}
\bibinfo{author}{\bibfnamefont{J.}~\bibnamefont{Salo}},
  \bibinfo{author}{\bibfnamefont{F.~W.~J.} \bibnamefont{Hekking}},
  \bibnamefont{and} \bibinfo{author}{\bibfnamefont{J.~P.} \bibnamefont{Pekola}}
  (\bibinfo{year}{2006}), \bibinfo{note}{[cond-mat/0605478]}.

\bibitem[{\citenamefont{Brosco et~al.}(2006)\citenamefont{Brosco, Fazio,
  Hekking, and Pekola}}]{brosco:024524}
\bibinfo{author}{\bibfnamefont{V.}~\bibnamefont{Brosco}},
  \bibinfo{author}{\bibfnamefont{R.}~\bibnamefont{Fazio}},
  \bibinfo{author}{\bibfnamefont{F.~W.~J.} \bibnamefont{Hekking}},
  \bibnamefont{and} \bibinfo{author}{\bibfnamefont{J.~P.}
  \bibnamefont{Pekola}}, \bibinfo{journal}{Phys. Rev. B}
  \textbf{\bibinfo{volume}{74}}, \bibinfo{eid}{024524} (\bibinfo{year}{2006}).

\bibitem[{\citenamefont{Heikkil\"a et~al.}(2004)\citenamefont{Heikkil\"a,
  Virtanen, Johansson, and Wilhelm}}]{heikkila:247005}
\bibinfo{author}{\bibfnamefont{T.~T.} \bibnamefont{Heikkil\"a}},
  \bibinfo{author}{\bibfnamefont{P.}~\bibnamefont{Virtanen}},
  \bibinfo{author}{\bibfnamefont{G.}~\bibnamefont{Johansson}},
  \bibnamefont{and} \bibinfo{author}{\bibfnamefont{F.~K.}
  \bibnamefont{Wilhelm}}, \bibinfo{journal}{Phys. Rev. Lett.}
  \textbf{\bibinfo{volume}{93}}, \bibinfo{eid}{247005} (\bibinfo{year}{2004}).

\bibitem[{\citenamefont{Ankerhold}(2006)}]{ankerholdup06}
\bibinfo{author}{\bibfnamefont{J.}~\bibnamefont{Ankerhold}}
  (\bibinfo{year}{2006}), \bibinfo{note}{[cond-mat/0607020]}.

\bibitem[{\citenamefont{Tobiska and Nazarov}(2004)}]{tobiska:106801}
\bibinfo{author}{\bibfnamefont{J.}~\bibnamefont{Tobiska}} \bibnamefont{and}
  \bibinfo{author}{\bibfnamefont{Y.~V.} \bibnamefont{Nazarov}},
  \bibinfo{journal}{Phys. Rev. Lett.} \textbf{\bibinfo{volume}{93}},
  \bibinfo{eid}{106801} (\bibinfo{year}{2004}).

\bibitem[{\citenamefont{Sonin}(2004)}]{sonin:140506}
\bibinfo{author}{\bibfnamefont{E.~B.} \bibnamefont{Sonin}},
  \bibinfo{journal}{Phys. Rev. B} \textbf{\bibinfo{volume}{70}},
  \bibinfo{eid}{140506(R)} (\bibinfo{year}{2004}).

\bibitem[{\citenamefont{Pekola}(2004)}]{pekola:206601}
\bibinfo{author}{\bibfnamefont{J.~P.} \bibnamefont{Pekola}},
  \bibinfo{journal}{Phys. Rev. Lett.} \textbf{\bibinfo{volume}{93}},
  \bibinfo{eid}{206601} (\bibinfo{year}{2004}).

\bibitem[{\citenamefont{Ankerhold and Grabert}(2005)}]{ankerhold:186601}
\bibinfo{author}{\bibfnamefont{J.}~\bibnamefont{Ankerhold}} \bibnamefont{and}
  \bibinfo{author}{\bibfnamefont{H.}~\bibnamefont{Grabert}},
  \bibinfo{journal}{Phys. Rev. Lett.} \textbf{\bibinfo{volume}{95}},
  \bibinfo{eid}{186601} (\bibinfo{year}{2005}).

\bibitem[{\citenamefont{Lindell et~al.}(2004)\citenamefont{Lindell, Delahaye,
  Sillanp\"a\"a, Heikkil\"a, Sonin, and Hakonen}}]{lindell:197002}
\bibinfo{author}{\bibfnamefont{R.~K.} \bibnamefont{Lindell}},
  \bibinfo{author}{\bibfnamefont{J.}~\bibnamefont{Delahaye}},
  \bibinfo{author}{\bibfnamefont{M.~A.} \bibnamefont{Sillanp\"a\"a}},
  \bibinfo{author}{\bibfnamefont{T.~T.} \bibnamefont{Heikkil\"a}},
  \bibinfo{author}{\bibfnamefont{E.~B.} \bibnamefont{Sonin}}, \bibnamefont{and}
  \bibinfo{author}{\bibfnamefont{P.~J.} \bibnamefont{Hakonen}},
  \bibinfo{journal}{Phys. Rev. Lett.} \textbf{\bibinfo{volume}{93}},
  \bibinfo{eid}{197002} (\bibinfo{year}{2004}).

\bibitem[{\citenamefont{Gustavsson
  et~al.}(2006{\natexlab{a}})\citenamefont{Gustavsson, Leturcq, Simovic,
  Schleser, Ihn, Studerus, Ensslin, Driscoll, and Gossard}}]{gustavsson:076605}
\bibinfo{author}{\bibfnamefont{S.}~\bibnamefont{Gustavsson}},
  \bibinfo{author}{\bibfnamefont{R.}~\bibnamefont{Leturcq}},
  \bibinfo{author}{\bibfnamefont{B.}~\bibnamefont{Simovic}},
  \bibinfo{author}{\bibfnamefont{R.}~\bibnamefont{Schleser}},
  \bibinfo{author}{\bibfnamefont{T.}~\bibnamefont{Ihn}},
  \bibinfo{author}{\bibfnamefont{P.}~\bibnamefont{Studerus}},
  \bibinfo{author}{\bibfnamefont{K.}~\bibnamefont{Ensslin}},
  \bibinfo{author}{\bibfnamefont{D.~C.} \bibnamefont{Driscoll}},
  \bibnamefont{and} \bibinfo{author}{\bibfnamefont{A.~C.}
  \bibnamefont{Gossard}}, \bibinfo{journal}{Phys. Rev. Lett.}
  \textbf{\bibinfo{volume}{96}}, \bibinfo{eid}{076605}
  (\bibinfo{year}{2006}{\natexlab{a}}).

\bibitem[{\citenamefont{Gustavsson
  et~al.}(2006{\natexlab{b}})\citenamefont{Gustavsson, Leturcq, Simovic,
  Schleser, Studerus, Ihn, Ensslin, D.C.Driscoll, and
  Gossard}}]{gustavssonup06a}
\bibinfo{author}{\bibfnamefont{S.}~\bibnamefont{Gustavsson}},
  \bibinfo{author}{\bibfnamefont{R.}~\bibnamefont{Leturcq}},
  \bibinfo{author}{\bibfnamefont{B.}~\bibnamefont{Simovic}},
  \bibinfo{author}{\bibfnamefont{R.}~\bibnamefont{Schleser}},
  \bibinfo{author}{\bibfnamefont{P.}~\bibnamefont{Studerus}},
  \bibinfo{author}{\bibfnamefont{T.}~\bibnamefont{Ihn}},
  \bibinfo{author}{\bibfnamefont{K.}~\bibnamefont{Ensslin}},
  \bibinfo{author}{\bibnamefont{D.C.Driscoll}}, \bibnamefont{and}
  \bibinfo{author}{\bibfnamefont{A.}~\bibnamefont{Gossard}}
  (\bibinfo{year}{2006}{\natexlab{b}}), \bibinfo{note}{[cond-mat/0605365]}.

\bibitem[{\citenamefont{Gustavsson
  et~al.}(2006{\natexlab{c}})\citenamefont{Gustavsson, Leturcq, Ihn, Ensslin,
  Reinwald, and Wegscheider}}]{gustavssonup06b}
\bibinfo{author}{\bibfnamefont{S.}~\bibnamefont{Gustavsson}},
  \bibinfo{author}{\bibfnamefont{R.}~\bibnamefont{Leturcq}},
  \bibinfo{author}{\bibfnamefont{T.}~\bibnamefont{Ihn}},
  \bibinfo{author}{\bibfnamefont{K.}~\bibnamefont{Ensslin}},
  \bibinfo{author}{\bibfnamefont{M.}~\bibnamefont{Reinwald}}, \bibnamefont{and}
  \bibinfo{author}{\bibfnamefont{W.}~\bibnamefont{Wegscheider}}
  (\bibinfo{year}{2006}{\natexlab{c}}), \bibinfo{note}{[cond-mat/0607192]}.

\bibitem[{\citenamefont{Caldeira and Leggett}(1983)}]{caldeira:374}
\bibinfo{author}{\bibfnamefont{A.~O.} \bibnamefont{Caldeira}} \bibnamefont{and}
  \bibinfo{author}{\bibfnamefont{A.~J.} \bibnamefont{Leggett}},
  \bibinfo{journal}{Ann. Phys.} \textbf{\bibinfo{volume}{149}},
  \bibinfo{pages}{374} (\bibinfo{year}{1983}).

\bibitem[{vol()}]{voltagenote}
\bibinfo{note}{One could equally well describe voltage fluctuations, but then
  the spectrum should be calculated for them.}

\bibitem[{fou()}]{fouriernote}
\bibinfo{note}{Note that this differs from the one used in
  Ref.~\onlinecite{ojanen:020501} by $4\pi^2$.}

\bibitem[{\citenamefont{Schoelkopf et~al.}(2003)\citenamefont{Schoelkopf,
  Clerk, Girvin, Lehnert, and Devoret}}]{schoelkopf02}
\bibinfo{author}{\bibfnamefont{R.}~\bibnamefont{Schoelkopf}},
  \bibinfo{author}{\bibfnamefont{A.}~\bibnamefont{Clerk}},
  \bibinfo{author}{\bibfnamefont{S.}~\bibnamefont{Girvin}},
  \bibinfo{author}{\bibfnamefont{K.}~\bibnamefont{Lehnert}}, \bibnamefont{and}
  \bibinfo{author}{\bibfnamefont{M.}~\bibnamefont{Devoret}},
  \emph{\bibinfo{title}{Quantum Noise in Mesoscopic Physics}}
  (\bibinfo{publisher}{Kluwer}, \bibinfo{address}{Dodrecht},
  \bibinfo{year}{2003}), vol.~\bibinfo{volume}{97} of
  \emph{\bibinfo{series}{Nato Science Series II}}.

\bibitem[{\citenamefont{Aguado and Kouwenhoven}(2000)}]{aguado:1986}
\bibinfo{author}{\bibfnamefont{R.}~\bibnamefont{Aguado}} \bibnamefont{and}
  \bibinfo{author}{\bibfnamefont{L.~P.} \bibnamefont{Kouwenhoven}},
  \bibinfo{journal}{Phys. Rev. Lett.} \textbf{\bibinfo{volume}{84}},
  \bibinfo{pages}{1986} (\bibinfo{year}{2000}).

\bibitem[{shp()}]{shpfouriernote}
\bibinfo{note}{Note that the Fourier convention in Eq.~\eqref{eq:third}
  slightly differs from Eq.~(42) in Ref.~\onlinecite{saloup06}.}

\bibitem[{\citenamefont{Blanter and B\"uttiker}(2000)}]{blanter:1}
\bibinfo{author}{\bibfnamefont{Y.}~\bibnamefont{Blanter}} \bibnamefont{and}
  \bibinfo{author}{\bibfnamefont{M.}~\bibnamefont{B\"uttiker}},
  \bibinfo{journal}{Phys. Rep.} \textbf{\bibinfo{volume}{336}},
  \bibinfo{pages}{1} (\bibinfo{year}{2000}).

\bibitem[{\citenamefont{Virtanen and Heikkil\"{a}}(2006)}]{virtanen:50}
\bibinfo{author}{\bibfnamefont{P.}~\bibnamefont{Virtanen}} \bibnamefont{and}
  \bibinfo{author}{\bibfnamefont{T.~T.} \bibnamefont{Heikkil\"{a}}},
  \bibinfo{journal}{New J. Phys.} \textbf{\bibinfo{volume}{8}},
  \bibinfo{pages}{50} (\bibinfo{year}{2006}).

\bibitem[{\citenamefont{Beenakker et~al.}(2003)\citenamefont{Beenakker,
  Kindermann, and Nazarov}}]{beenakker:176802}
\bibinfo{author}{\bibfnamefont{C.~W.~J.} \bibnamefont{Beenakker}},
  \bibinfo{author}{\bibfnamefont{M.}~\bibnamefont{Kindermann}},
  \bibnamefont{and} \bibinfo{author}{\bibfnamefont{Y.~V.}
  \bibnamefont{Nazarov}}, \bibinfo{journal}{Phys. Rev. Lett.}
  \textbf{\bibinfo{volume}{90}}, \bibinfo{eid}{176802} (\bibinfo{year}{2003}).

\bibitem[{\citenamefont{Schep and Bauer}(1997)}]{schep:3015}
\bibinfo{author}{\bibfnamefont{K.~M.} \bibnamefont{Schep}} \bibnamefont{and}
  \bibinfo{author}{\bibfnamefont{G.~E.~W.} \bibnamefont{Bauer}},
  \bibinfo{journal}{Phys. Rev. Lett.} \textbf{\bibinfo{volume}{78}},
  \bibinfo{pages}{3015} (\bibinfo{year}{1997}).

\bibitem[{\citenamefont{Nazarov}(1994)}]{nazarov:134}
\bibinfo{author}{\bibfnamefont{Y.~V.} \bibnamefont{Nazarov}},
  \bibinfo{journal}{Phys. Rev. Lett.} \textbf{\bibinfo{volume}{73}},
  \bibinfo{pages}{134} (\bibinfo{year}{1994}).

\bibitem[{\citenamefont{Baranger and Mello}(1994)}]{baranger:142}
\bibinfo{author}{\bibfnamefont{H.~U.} \bibnamefont{Baranger}} \bibnamefont{and}
  \bibinfo{author}{\bibfnamefont{P.~A.} \bibnamefont{Mello}},
  \bibinfo{journal}{Phys. Rev. Lett.} \textbf{\bibinfo{volume}{73}},
  \bibinfo{pages}{142} (\bibinfo{year}{1994}).

\bibitem[{rel()}]{reldirectionnote}
\bibinfo{note}{Here we assume that in the absence of the noise source, the
  qubit relaxation is independent of the direction of the fields.}

\bibitem[{\citenamefont{Pekola et~al.}(2005)\citenamefont{Pekola, Nieminen,
  Meschke, Kivioja, Niskanen, and Vartiainen}}]{pekola:197004}
\bibinfo{author}{\bibfnamefont{J.~P.} \bibnamefont{Pekola}},
  \bibinfo{author}{\bibfnamefont{T.~E.} \bibnamefont{Nieminen}},
  \bibinfo{author}{\bibfnamefont{M.}~\bibnamefont{Meschke}},
  \bibinfo{author}{\bibfnamefont{J.~M.} \bibnamefont{Kivioja}},
  \bibinfo{author}{\bibfnamefont{A.~O.} \bibnamefont{Niskanen}},
  \bibnamefont{and} \bibinfo{author}{\bibfnamefont{J.~J.}
  \bibnamefont{Vartiainen}}, \bibinfo{journal}{Phys. Rev. Lett.}
  \textbf{\bibinfo{volume}{95}}, \bibinfo{eid}{197004} (\bibinfo{year}{2005}).

%\bibitem[{\citenamefont{Ankerhold}()}]{ankerholdpc}
%\bibinfo{author}{\bibfnamefont{J.}~\bibnamefont{Ankerhold}},
%  \bibinfo{note}{private communication}.

\bibitem[{sec()}]{secondterm}
\bibinfo{note}{To be precise, we assume here that the inverse mass is perturbed
  by the current fluctuations. In some cases, the next-order term of the form
  $(\delta m)^2/m^3$ would also contribute in the third order in $\delta
  m(\delta I)$.}

\bibitem[{\citenamefont{Majer et~al.}(2005)\citenamefont{Majer, Paauw, ter
  Haar, Harmans, and Mooij}}]{majer:090501}
\bibinfo{author}{\bibfnamefont{J.~B.} \bibnamefont{Majer}},
  \bibinfo{author}{\bibfnamefont{F.~G.} \bibnamefont{Paauw}},
  \bibinfo{author}{\bibfnamefont{A.~C.~J.} \bibnamefont{ter Haar}},
  \bibinfo{author}{\bibfnamefont{C.~J. P.~M.} \bibnamefont{Harmans}},
  \bibnamefont{and} \bibinfo{author}{\bibfnamefont{J.~E.} \bibnamefont{Mooij}},
  \bibinfo{journal}{Phys. Rev. Lett.} \textbf{\bibinfo{volume}{94}},
  \bibinfo{eid}{090501} (\bibinfo{year}{2005}).

\bibitem[{\citenamefont{Chiorescu et~al.}(2003)\citenamefont{Chiorescu,
  Nakamura, Harmans, and Mooij}}]{chiorescu:1869}
\bibinfo{author}{\bibfnamefont{I.}~\bibnamefont{Chiorescu}},
  \bibinfo{author}{\bibfnamefont{Y.}~\bibnamefont{Nakamura}},
  \bibinfo{author}{\bibfnamefont{C.~J. P.~M.} \bibnamefont{Harmans}},
  \bibnamefont{and} \bibinfo{author}{\bibfnamefont{J.~E.} \bibnamefont{Mooij}},
  \bibinfo{journal}{Science} \textbf{\bibinfo{volume}{299}},
  \bibinfo{pages}{1869} (\bibinfo{year}{2003}).

\bibitem[{\citenamefont{Martinis et~al.}(2003)\citenamefont{Martinis, Nam,
  Aumentado, Lang, and Urbina}}]{martinis:094510}
\bibinfo{author}{\bibfnamefont{J.~M.} \bibnamefont{Martinis}},
  \bibinfo{author}{\bibfnamefont{S.}~\bibnamefont{Nam}},
  \bibinfo{author}{\bibfnamefont{J.}~\bibnamefont{Aumentado}},
  \bibinfo{author}{\bibfnamefont{K.~M.} \bibnamefont{Lang}}, \bibnamefont{and}
  \bibinfo{author}{\bibfnamefont{C.}~\bibnamefont{Urbina}},
  \bibinfo{journal}{Phys. Rev. B} \textbf{\bibinfo{volume}{67}},
  \bibinfo{eid}{094510} (\bibinfo{year}{2003}).

\bibitem[{\citenamefont{Martinis et~al.}(2002)\citenamefont{Martinis, Nam,
  Aumentado, and Urbina}}]{martinis:117901}
\bibinfo{author}{\bibfnamefont{J.~M.} \bibnamefont{Martinis}},
  \bibinfo{author}{\bibfnamefont{S.}~\bibnamefont{Nam}},
  \bibinfo{author}{\bibfnamefont{J.}~\bibnamefont{Aumentado}},
  \bibnamefont{and} \bibinfo{author}{\bibfnamefont{C.}~\bibnamefont{Urbina}},
  \bibinfo{journal}{Phys. Rev. Lett.} \textbf{\bibinfo{volume}{89}},
  \bibinfo{pages}{117901} (\bibinfo{year}{2002}).

\bibitem[{\citenamefont{Nakamura et~al.}(1999)\citenamefont{Nakamura, Pashkin,
  and Tsai}}]{nakamura:786}
\bibinfo{author}{\bibfnamefont{Y.}~\bibnamefont{Nakamura}},
  \bibinfo{author}{\bibfnamefont{Y.~A.} \bibnamefont{Pashkin}},
  \bibnamefont{and} \bibinfo{author}{\bibfnamefont{J.~S.} \bibnamefont{Tsai}},
  \bibinfo{journal}{Nature} \textbf{\bibinfo{volume}{398}},
  \bibinfo{pages}{786} (\bibinfo{year}{1999}).

\bibitem[{\citenamefont{Nagaev et~al.}(2004)\citenamefont{Nagaev, Pilgram, and
  Buttiker}}]{nagaev:176804}
\bibinfo{author}{\bibfnamefont{K.~E.} \bibnamefont{Nagaev}},
  \bibinfo{author}{\bibfnamefont{S.}~\bibnamefont{Pilgram}}, \bibnamefont{and}
  \bibinfo{author}{\bibfnamefont{M.}~\bibnamefont{Buttiker}},
  \bibinfo{journal}{Phys. Rev. Lett.} \textbf{\bibinfo{volume}{92}},
  \bibinfo{eid}{176804} (\bibinfo{year}{2004}).

\bibitem[{\citenamefont{Hekking and Pekola}(2006)}]{hekking:056603}
\bibinfo{author}{\bibfnamefont{F.~W.~J.} \bibnamefont{Hekking}}
  \bibnamefont{and} \bibinfo{author}{\bibfnamefont{J.~P.}
  \bibnamefont{Pekola}}, \bibinfo{journal}{Phys. Rev. Lett.}
  \textbf{\bibinfo{volume}{96}}, \bibinfo{eid}{056603} (\bibinfo{year}{2006}).

\bibitem[{\citenamefont{Pilgram et~al.}(2004)\citenamefont{Pilgram, Nagaev, and
  Buttiker}}]{pilgram:045304}
\bibinfo{author}{\bibfnamefont{S.}~\bibnamefont{Pilgram}},
  \bibinfo{author}{\bibfnamefont{K.~E.} \bibnamefont{Nagaev}},
  \bibnamefont{and} \bibinfo{author}{\bibfnamefont{M.}~\bibnamefont{Buttiker}},
  \bibinfo{journal}{Phys. Rev. B} \textbf{\bibinfo{volume}{70}},
  \bibinfo{eid}{045304} (\bibinfo{year}{2004}).

\bibitem[{\citenamefont{Makhlin et~al.}(2001)\citenamefont{Makhlin, Sch\"on,
  and Shnirman}}]{makhlin:357}
\bibinfo{author}{\bibfnamefont{Y.}~\bibnamefont{Makhlin}},
  \bibinfo{author}{\bibfnamefont{G.}~\bibnamefont{Sch\"on}}, \bibnamefont{and}
  \bibinfo{author}{\bibfnamefont{A.}~\bibnamefont{Shnirman}},
  \bibinfo{journal}{Rev. Mod. Phys.} \textbf{\bibinfo{volume}{73}},
  \bibinfo{pages}{357} (\bibinfo{year}{2001}).

\bibitem[{\citenamefont{Vion et~al.}(2002)\citenamefont{Vion, Aassime, Cottet,
  Joyez, Pothier, Urbina, Esteve, and Devoret}}]{vion:886}
\bibinfo{author}{\bibfnamefont{D.}~\bibnamefont{Vion}},
  \bibinfo{author}{\bibfnamefont{A.}~\bibnamefont{Aassime}},
  \bibinfo{author}{\bibfnamefont{A.}~\bibnamefont{Cottet}},
  \bibinfo{author}{\bibfnamefont{P.}~\bibnamefont{Joyez}},
  \bibinfo{author}{\bibfnamefont{H.}~\bibnamefont{Pothier}},
  \bibinfo{author}{\bibfnamefont{C.}~\bibnamefont{Urbina}},
  \bibinfo{author}{\bibfnamefont{D.}~\bibnamefont{Esteve}}, \bibnamefont{and}
  \bibinfo{author}{\bibfnamefont{M.~H.} \bibnamefont{Devoret}},
  \bibinfo{journal}{Science} \textbf{\bibinfo{volume}{296}},
  \bibinfo{pages}{886} (\bibinfo{year}{2002}).

\bibitem[{\citenamefont{Roschier et~al.}(2005)\citenamefont{Roschier,
  Sillanp\"a\"a, and Hakonen}}]{roschier:024530}
\bibinfo{author}{\bibfnamefont{L.}~\bibnamefont{Roschier}},
  \bibinfo{author}{\bibfnamefont{M.}~\bibnamefont{Sillanp\"a\"a}},
  \bibnamefont{and} \bibinfo{author}{\bibfnamefont{P.}~\bibnamefont{Hakonen}},
  \bibinfo{journal}{Phys. Rev. B} \textbf{\bibinfo{volume}{71}},
  \bibinfo{eid}{024530} (\bibinfo{year}{2005}).

\bibitem[{\citenamefont{Sillanp\"a\"a et~al.}(2005)\citenamefont{Sillanp\"a\"a,
  Lehtinen, Paila, Makhlin, Roschier, and Hakonen}}]{sillanpaa:206806}
\bibinfo{author}{\bibfnamefont{M.~A.} \bibnamefont{Sillanp\"a\"a}},
  \bibinfo{author}{\bibfnamefont{T.}~\bibnamefont{Lehtinen}},
  \bibinfo{author}{\bibfnamefont{A.}~\bibnamefont{Paila}},
  \bibinfo{author}{\bibfnamefont{Y.}~\bibnamefont{Makhlin}},
  \bibinfo{author}{\bibfnamefont{L.}~\bibnamefont{Roschier}}, \bibnamefont{and}
  \bibinfo{author}{\bibfnamefont{P.~J.} \bibnamefont{Hakonen}},
  \bibinfo{journal}{Phys. Rev. Lett.} \textbf{\bibinfo{volume}{95}},
  \bibinfo{eid}{206806} (\bibinfo{year}{2005}).

\bibitem[{\citenamefont{Duty et~al.}(2005)\citenamefont{Duty, Johansson, Bladh,
  Gunnarsson, Wilson, and Delsing}}]{duty:206807}
\bibinfo{author}{\bibfnamefont{T.}~\bibnamefont{Duty}},
  \bibinfo{author}{\bibfnamefont{G.}~\bibnamefont{Johansson}},
  \bibinfo{author}{\bibfnamefont{K.}~\bibnamefont{Bladh}},
  \bibinfo{author}{\bibfnamefont{D.}~\bibnamefont{Gunnarsson}},
  \bibinfo{author}{\bibfnamefont{C.}~\bibnamefont{Wilson}}, \bibnamefont{and}
  \bibinfo{author}{\bibfnamefont{P.}~\bibnamefont{Delsing}},
  \bibinfo{journal}{Phys. Rev. Lett.} \textbf{\bibinfo{volume}{95}},
  \bibinfo{eid}{206807} (\bibinfo{year}{2005}).

\bibitem[{jjs()}]{jjsecondterm}
\bibinfo{note}{The criterion for neglecting the term of the form $(\delta
  L)^2/(L^2)$ (see Ref.~\onlinecite{secondterm}) here is $\sin(2\pi
  \Phi_x/\Phi_0) \gg \cos(2\pi \Phi_x/\Phi_0)$.}

\bibitem[{\citenamefont{Lesovik and Chtchelkatchev}(2003)}]{lesovik:393}
\bibinfo{author}{\bibfnamefont{G.~B.} \bibnamefont{Lesovik}} \bibnamefont{and}
  \bibinfo{author}{\bibfnamefont{N.~M.} \bibnamefont{Chtchelkatchev}},
  \bibinfo{journal}{JETP Lett.} \textbf{\bibinfo{volume}{77}},
  \bibinfo{pages}{393} (\bibinfo{year}{2003}).

\end{thebibliography}
\end{document}